\documentclass[fleqn,usenatbib]{mnras}
\usepackage{general}
\usepackage{mathptmx}
\usepackage{txfonts}
\usepackage[T1]{fontenc}
\usepackage{ae,aecompl}
\title[Formation of peptide-like molecules]{The formation of peptide-like molecules on interstellar dust grains}

\author[N. F. W. Ligterink et al.]{
N. F. W. Ligterink,$^{1,2}$\thanks{E-mail: ligterink@strw.leidenuniv.nl}
J. Terwisscha van Scheltinga,$^{1,2}$ 
V. Taquet,$^{3}$ \and \
J. K. J{\o}rgensen,$^{4}$ 
S. Cazaux,$^{5}$
E. F. van Dishoeck$^{2,6}$
and H. Linnartz$^{1}$
\\
$^{1}$Raymond and Beverly Sackler Laboratory for Astrophysics, Leiden Observatory, Leiden\\ University, PO Box 9513, 2300 RA Leiden, The Netherlands\\      
$^{2}$Leiden Observatory, Leiden University, PO Box 9513, 2300 RA Leiden, The Netherlands\\
$^{3}$INAF-Osservatorio Astrofisico di Arcetri, Largo E. Fermi 5, I-50125, Florence, Italy\\
$^{4}$Centre for Star and Planet Formation, Niels Bohr Institute \& Natural History Museum of Denmark, University of Copenhagen,\\ \phantom{'}{\O}ster Voldgade 5--7, 1350 Copenhagen K., Denmark\\
$^{5}$Faculty of Aerospace Engineering, Delft University of Technology, Delft, Netherlands\\
$^{6}$ Max-Planck Institut f\"{u}r Extraterrestrische Physik (MPE), Giessenbachstr. 1, 85748 Garching, Germany\\
}

\date{Accepted XXX. Received YYY; in original form ZZZ}

\pubyear{2018}

\hypersetup{draft}
\begin{document}
\def\nratio{\ensuremath{\foun/\fifn}} 
\def\cratio{\ensuremath{\rm ^{12}C/^{13}C}}
\label{firstpage}
\pagerange{\pageref{firstpage}--\pageref{lastpage}}
\maketitle

\begin{abstract}
Molecules with an amide functional group resemble peptide bonds, the molecular bridges that connect amino acids, and may thus be relevant in processes that lead to the formation of life. In this study, the solid state formation of some of the smallest amides is investigated in the laboratory. To this end, CH$_{4}$:HNCO ice mixtures at 20~K are irradiated with far-UV photons, where the radiation is used as a tool to produce the radicals required for the formation of the amides. Products are identified and investigated with infrared spectroscopy and temperature programmed desorption mass spectrometry. 

The laboratory data show that NH$_{2}$CHO, CH$_{3}$NCO, NH$_{2}$C(O)NH$_{2}$, CH$_{3}$C(O)NH$_{2}$ and CH$_{3}$NH$_{2}$ can simultaneously be formed. The NH$_{2}$CO radical is found to be key in the formation of larger amides. In parallel, ALMA observations towards the low-mass protostar IRAS~16293--2422B are analysed in search of CH$_{3}$NHCHO (N-methylformamide) and CH$_{3}$C(O)NH$_{2}$ (acetamide). CH$_{3}$C(O)NH$_{2}$ is tentatively detected towards IRAS~16293--2422B at an abundance comparable with those found towards high-mass sources. The combined laboratory and observational data indicates that NH$_{2}$CHO and CH$_{3}$C(O)NH$_{2}$ are chemically linked and form in the ice mantles of interstellar dust grains. A solid-state reaction network for the formation of these amides is proposed.

\end{abstract}




\section{Introduction}
\label{sec.int}

\begin{figure*}
\begin{center}
\includegraphics[width=\hsize]{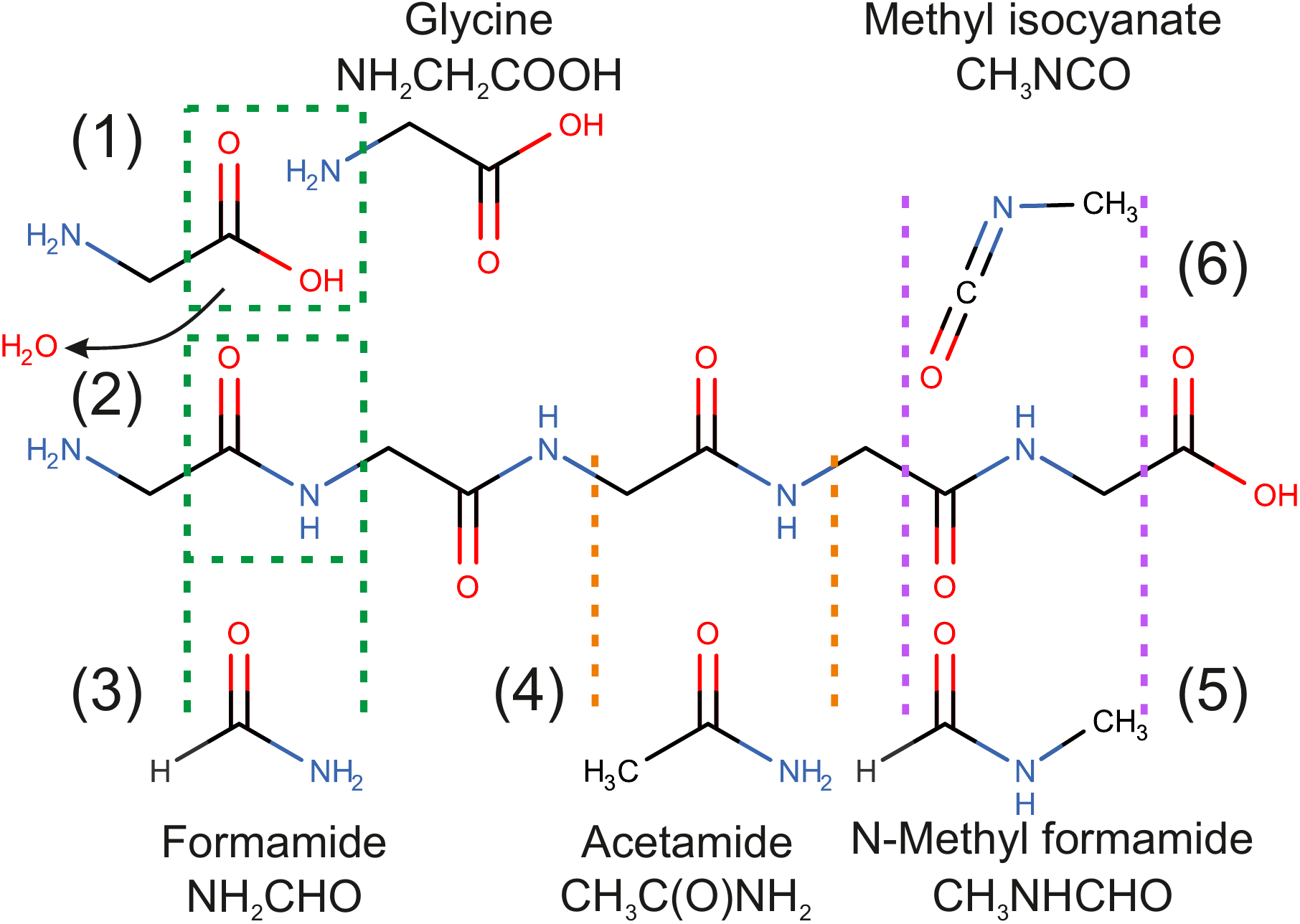}
\caption{The reaction between the acid and base groups of two glycine molecules (1) forms a peptide bonded molecular chain (2). The peptide bond shows similarities to the smallest amide formamide (3), but the larger peptide chain incorporates structures that are similar to acetamide (4), N-methylformamide (5) and methyl isocyanate (6). Note that carbon atoms are not indicated, except for terminal groups.}
\label{fig.pep_bond}
\end{center}
\end{figure*}

Prebiotic molecules are species that resemble functional groups of biogenic molecules and are thought to be involved in the formation of molecules that are relevant to life, such as amino acids, nucleobases and sugars \citep{herbstdishoeck2009,caselliceccarelli2012}. The interstellar presence of prebiotic molecules supports the idea that the building blocks of life may have an extraterrestrial origin. A number of these molecules have been detected in the InterStellar Medium (ISM), such as the simplest ``sugar'' glycolaldehyde \citep[CH(O)CH$_{2}$OH,][]{hollis2004,jorgensen2012,jorgensen2016} and precursor molecules to the amino acid glycine, such as methylamine \citep[CH$_{3}$NH$_{2}$,][]{kaifu1974} and aminoacetonitril \citep[NH$_{2}$CH$_{2}$CN,][]{belloche2008}. Among prebiotics, molecules with an amide (--NH--C(O)--) or amide-like structure, such as isocyanic acid (HNCO), hereafter generally called amides, are of particular interest because they resemble a peptide bond, see Fig \ref{fig.pep_bond}. In terrestrial biochemistry amino acids are connected by peptide bonds resulting in long chains which eventually form proteins, the engines of life. Reactions involving molecules with an amide functional group offer alternative pathways to form peptide chains.

Amides are widespread throughout the ISM. HNCO and formamide (NH$_{2}$CHO) are the most abundant ones and have been detected in a large variety of interstellar sources \citep[e.g.][]{bisschop2007,kahane2013,adande2013,corby2015,bergner2017} and comets, including 67P/Churyumov-Gerasimenko \citep[67P/C-G,][]{bockelee-morvan2000,goesmann2015,altwegg2017}. Observational evidence exists for a chemical relationship between HNCO and NH$_{2}$CHO, which is thought to originate in interstellar ice \citep{bisschop2007,lopez-sepulcre2015,coutens2016}. In the form of the OCN$^{-}$ anion, HNCO has been directly detected in interstellar ices at abundances as high as 2\% with respect to water \citep{lacy1984,gibb2004,vanbroekhuizen2005}. Tentatively, the presence of formamide in interstellar ice has been claimed towards NGC 7538 IRS9 \citep{raunier2004}. 

The more complex molecule acetamide (CH$_{3}$C(O)NH$_{2}$) has been detected towards Sagittarius B2 (Sgr B2) and Orion KL \citep{hollis2006,cernicharo2016,belloche2017} and on 67P/C-G \citep{goesmann2015,altwegg2017}. Formation of this molecule has been linked to that of formamide \citep{halfen2011}, although it is inconclusive whether gas-phase or solid-state chemistry is involved \citep[see also][]{quan2007}. Methyl isocyanate (CH$_{3}$NCO) has been detected towards Sgr B2 and Orion KL \citep{halfen2015,cernicharo2016} and recently towards the sun-like protostar IRAS 16293--2422 \citep{ligterink2017a,martin-domenech2017}. Its formational origin is likely found in interstellar ices, although some non-negligible gas-phase production routes are available \citep{quenard2018}. Hydrogenation of CH$_{3}$NCO is hypothesised to lead to N-methylformamide (CH$_{3}$NHCHO), a molecule that has tentatively been detected towards Sgr B2 \citep{belloche2017}. Carbamide, also known as urea (NH$_{2}$C(O)NH$_{2}$), has tentatively been identified towards Sgr B2 as well \citep{remijan2014}. Finally, cyanamide (NH$_{2}$CN) has been observed towards various galactic and extragalactic sources \citep[e.g.][]{turner1975,martin2006,coutens2017}.

The high interstellar abundances of HNCO and NH$_{2}$CHO have resulted in many solid-state laboratory studies with the aim to understand their formation \citep{hagen1979,gerakines2004,raunier2004,jones2011,islam2014,munozcaro2014,fedoseev2015,noble2015,fedoseev2016,kanuchova2016,fedoseev2018}. In these studies, ice mixtures containing a source of carbon, like CH$_{3}$OH or CO, and a source of nitrogen, such as HCN, HNCO, NH$_{3}$, N$_{2}$ or NO are hydrogenated and/or energetically processed. A number of mechanisms have been shown to produce these species, such as the NH + CO reaction to produce HNCO \citep{fedoseev2015} and the NH$_{2}$ + CHO radical combination to produce NH$_{2}$CHO \citep{jones2011}. \citet{raunier2004} proposed that HNCO can be hydrogenated to NH$_{2}$CHO by hot H-atom addition (i.e. hydrogen atoms produced by energetic dissociation processes that carry enough excess energy to overcome reaction barriers). On the other hand, \citet{noble2015} showed that hydrogenation of HNCO with ``cold'' ($\sim$300~K) hydrogen atoms produced in a beam line does not result in the formation of NH$_{2}$CHO. \\

The larger amides NH$_{2}$C(O)NH$_{2}$ and CH$_{3}$C(O)NH$_{2}$ have been produced in various ice experiments \citep[e.g.][]{berger1961,agarwal1985,bernstein1995,raunier2004,henderson2015,forstel2016}, but formation mechanisms have not been extensively investigated. Some reactions have been proposed, such as the NH$_{2}$ + NH$_{2}$CO radical addition to form NH$_{2}$C(O)NH$_{2}$ \citep{agarwal1985,raunier2004}. Modelling investigations have predicted the formation of CH$_{3}$C(O)NH$_{2}$ through the CH$_{3}$ + HNCO reaction followed by hydrogenation \citep{garrod2008} or the hydrogen abstraction of NH$_{2}$CHO followed by CH$_{3}$ addition \citep{belloche2017}. The formation of CH$_{3}$NHCHO has been claimed in far-UV (also known as vacuum-UV or V-UV) irradiated CH$_{3}$NH$_{2}$:CO ice mixtures through the reaction CH$_{3}$NH + CHO \citep{bossa2012}, while modelling investigations have shown that hydrogenation of CH$_{3}$NCO is one of the main channels of CH$_{3}$NHCHO formation \citep{belloche2017}. Recently, the solid-state reaction CH$_{3}$ + (H)NCO was proposed as the most likely candidate to explain the formation of CH$_{3}$NCO \citep{ligterink2017a}. Other solid-state pathways, such as formation via a HCN$^{...}$CO van der Waals complex, have also been proposed as relevant pathways in modelling studies \citep{majumdar2018}.

The aim of this work is to elucidate the chemical network that links various small amides that have been detected in the ISM and explain their formation. This work is complementary to that of \citet{ligterink2017a} on CH$_{3}$NCO and investigates reactions that can occur simultaneously with the formation of this molecule. Ice mixtures of CH$_{4}$:HNCO, two astronomically relevant precursor species, are irradiated with far-UV radiation. The far-UV radiation is used as a tool to form radicals, which engage in recombination reaction to form amides, amines and other molecules. On interstellar dust grains these radicals could be formed by far-UV photodissociation, but also non-energetically by hydrogenation of atomic carbon, oxygen and nitrogen.

This paper is organised in the following way. Section \ref{sec.lab} discusses the laboratory set-up and measurement protocol. The results of the experiments are presented in Sec. \ref{sec.res}. Observations and the comparison between laboratory and observational results are presented in Sec. \ref{sec.obs}, followed by the discussion in Sec. \ref{sec.dis}. The conclusions of this work are presented in Sec. \ref{sec.con}.

\begin{table}
     \caption[]{Overview of performed far-UV irradiation experiments on ice mixtures.}
         $$
         \begin{tabular}{l l l l l l}
            \hline
            \hline
            \noalign{\smallskip}
            Exp. & $N$(HNCO) & $N$(CH$_{4}$) & $N$(CO)$^{a}$ & Lyman-$\alpha$ \\
            & \multicolumn{3}{c}{ML (10$^{15}$ molecules cm$^{-2}$)}  & High/Low\\
            \noalign{\smallskip}
            \hline
            \noalign{\smallskip}
			1 & 14.3 & 17.0 & -- & H \\          	
          	2 & 17.4 & - & -- & H \\
          	3 & 15.1 & 15.7$^{b}$ & -- & H \\
            \noalign{\smallskip}
            \hline
            \noalign{\smallskip}       	
          	4 & 29.9 & 5.5 & -- & H  \\
          	5 & 11.4 & 24.6 & -- & L \\
          	\hline
          	6 & 2.9 & 8.5 & 95.6 & L & \\
          	7 & 4.8 & 13.9 & 164.2 & L & \\

            \noalign{\smallskip}
            \hline
            
         \end{tabular}     
         $$     
	\emph{\rm Notes. $^{a}$Total $^{12+13}$CO column density calculated from the $^{13}$CO band multiplied by 91. $^{b}$Experiment using $^{13}$CH$_{4}$, the bandstrength value of 1.1$\times$10$^{-17}$ cm molecule$^{-1}$ is assumed to apply to the $^{13}$CH$_{4}$ degenerate stretching mode as well. Other bandstrength values are found in Table~\ref{tab.IRfeat}.} 
	\label{tab.exper}
\end{table}

\begin{table*}
     \caption[]{Peak positions and transmission bandstrengths of precursor and product species}
         \label{tab.IR_param}
         
         \begin{tabular}{l l l l l l}
            \hline
            \hline
            \noalign{\smallskip}
            Species & Name & band & \multicolumn{2}{c}{Peak position} & Bandstrength \\
            		&      &	   & \multicolumn{2}{c}{(cm$^{-1}$)} & cm molecule$^{-1}$ \\
            		&      &	   & Literature & Experiment* & Transmission \\
            \noalign{\smallskip}
            \hline
            \noalign{\smallskip}
			HNCO (water poor) & Isocyanic acid & OCN str.$^{a}$ & 2260 & 2266 & 7.8$\times$10$^{-17}$ \\
			CH$_{4}$ & Methane & d-str.$^{b}$ & 1301 & 1302 & 7.3$\times$10$^{-18}$ \\
			CH$_{4}$ & Methane & d-str.$^{b}$ & 3010 & 3010 & 1.1$\times$10$^{-17}$ \\
			CO & Carbon monoxide & CO str.$^{c}$ & 2138 & 2142 & 1.1$\times$10$^{-17}$ \\
\noalign{\smallskip}
\hline
\noalign{\smallskip}
			OCN$^{-}$ (water poor) & Cyanate anion & OCN str.$^{a}$ & 2160 & 2170 & 1.3$\times$10$^{-16}$ \\
			CO$_{2}$ & Carbon dioxide & CO a-str.$^{c}$ & 2342 & 2341 & 7.6$\times$10$^{-17}$ \\
			HCN & Hydrogen cyanide & CN str.$^{d}$** & 2099 & 2108 & -- \\
			CH$_{3}$NCO & Methyl isocyanate & NCO a-str.$^{e}$** & 2322 & 2322 & -- \\
			CH$_{3}$CH$_{3}$ & Ethane & CH$_{3}$ d-str.$^{f}$ & 2975 & 2976 & 6.5$\times$10$^{-18}$ \\
			NH$_{4}^{+}$ & Ammonium cation & deform.$^{a}$ & 1485 & 1466 & 4.6$\times$10$^{-17}$ \\
			NH$_{2}$CHO & Formamide & CO str.$^{g,h}$ & 1700 & $\sim$1687 & 3.3$\times$10$^{-17}$ \\			
			NH$_{2}$CONH$_{2}$ & Carbamide & CO str.$^{h}$ & 1590 & -- & -- \\			
			NH$_{2}$CONH$_{2}$ & Carbamide & NH sym. bend$^{h}$ & 1675 & $\sim$1687 & -- \\
			NH$_{2}$CONH$_{2}$ & Carbamide & NH asym. bend$^{h}$ & 1630 & $\sim$1687 & -- \\
            \hline
            
         \end{tabular}     
  
	\emph{\rm Notes. *Peak positions found for experiment 1 (see Table~\ref{tab.exper}); **Indicates IR data obtained from reflection experiments. $^{a}$\citet{vanbroekhuizen2004}; $^{b}$\citet{hudgins1993,boogert1997}; $^{c}$\citet{bouilloud2015}; $^{d}$\citet{gerakines2004}; $^{e}$\citet{ligterink2017a}; $^{f}$\citet{gerakines1996}; $^{g}$\citet{wexler1967}; $^{h}$\citet{raunier2004}} 
	\label{tab.IRfeat}
\end{table*}

\begin{figure}
\begin{center}
\includegraphics[width=\hsize]{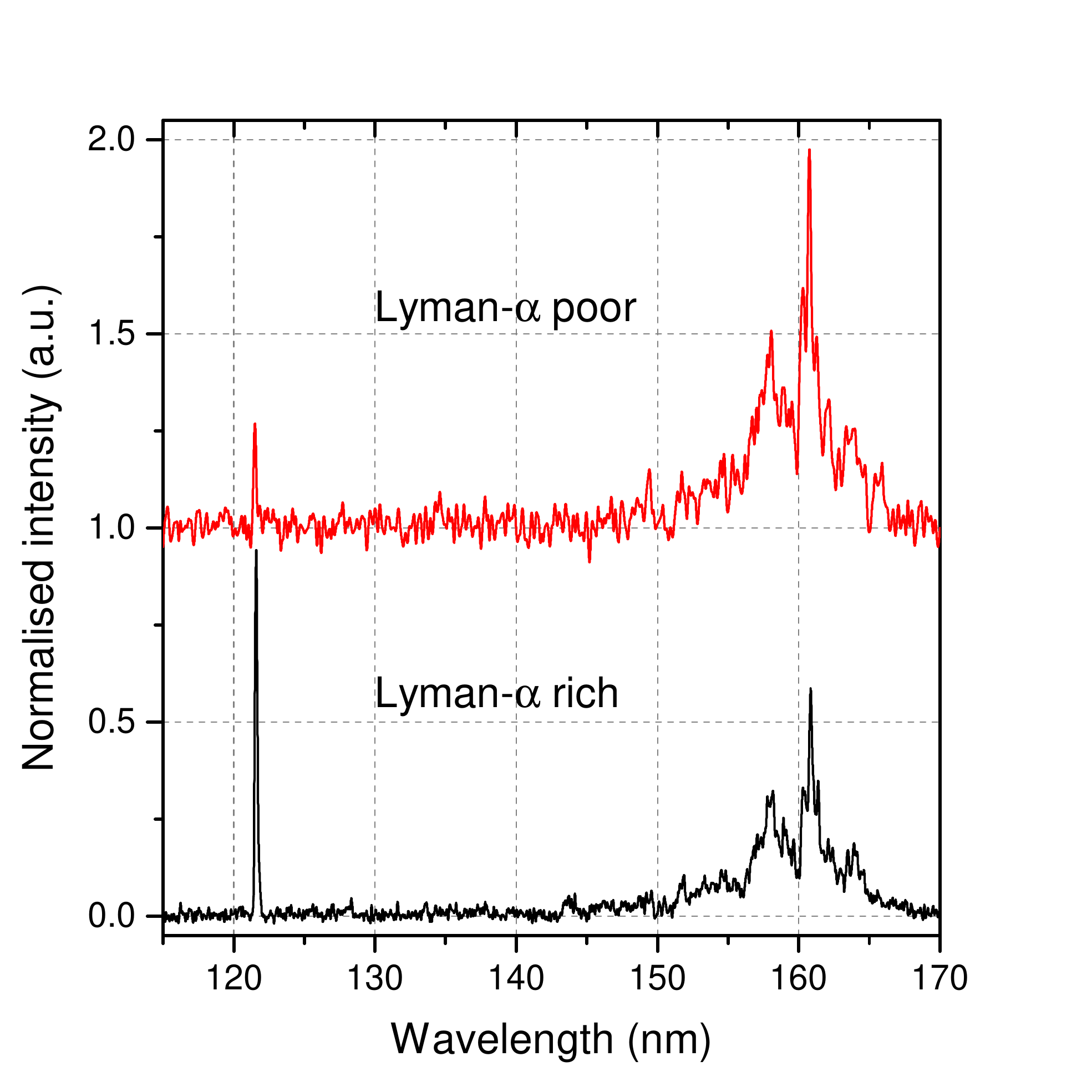}
\caption{Far-UV spectrum of the MDHL emission between 115 and 170 nm. The top spectrum (red) shows lamp emission poor in Lyman-$\alpha$, while the bottom spectrum (black) shows lamp emission rich in Lyman-$\alpha$.}
\label{fig.VUV_comp}
\end{center}
\end{figure}

\begin{figure}
\begin{center}
\includegraphics[width=\hsize]{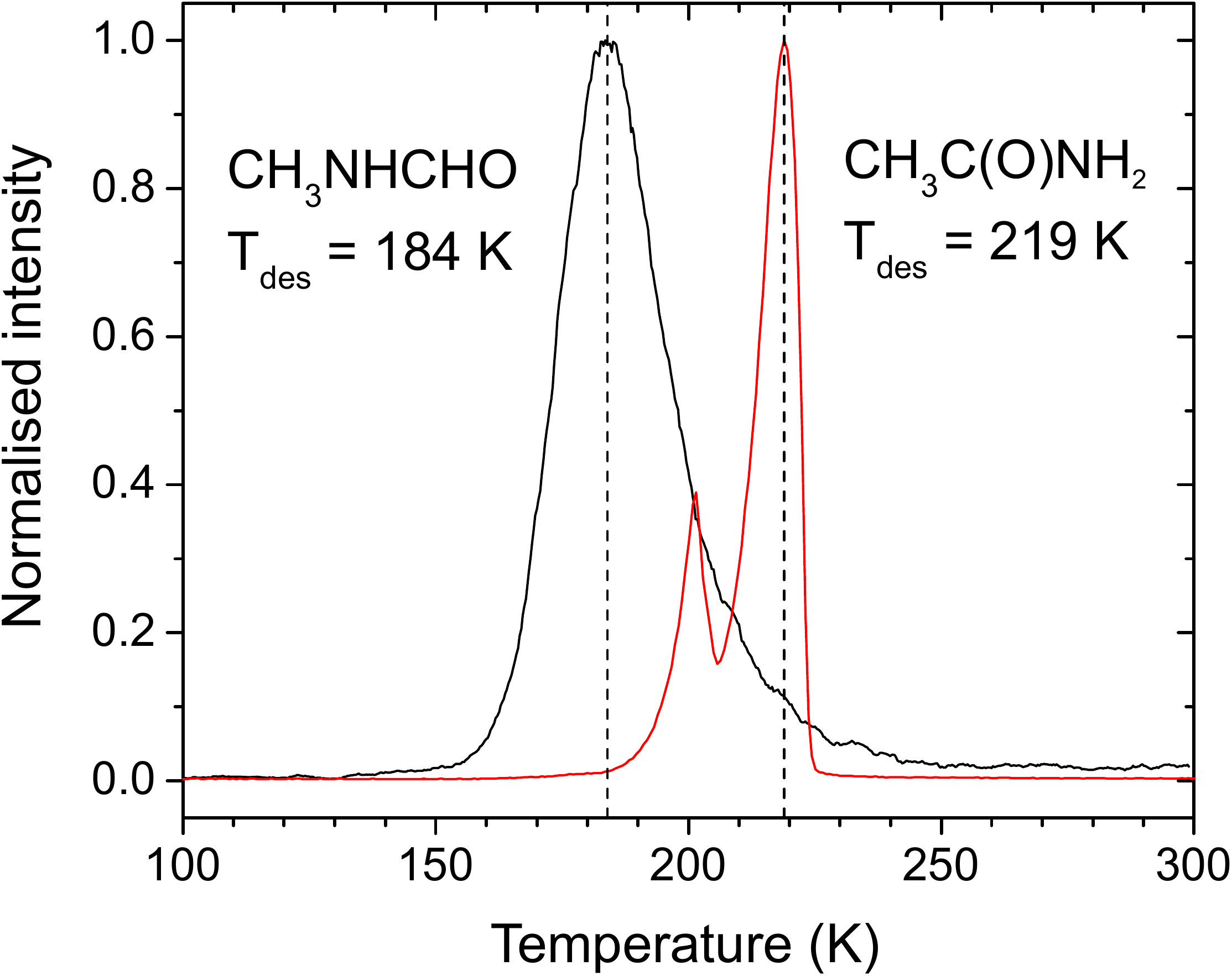}
\caption{TPD traces of $m/z$ 59 of pure CH$_{3}$NHCHO (black) and pure CH$_{3}$C(O)NH$_{2}$ (red). Note that CH$_{3}$C(O)NH$_{2}$ has a double desorption peak with the first peak potentially caused by a phase change in the ice.}
\label{fig.Tdes}
\end{center}
\end{figure} 

\section{Experimental}
\label{sec.lab}

\subsection{Set-up and protocol}

For this study the CryoPAD2 set-up in the Sackler Laboratory for Astrophysics is used, which has been described in \citet{ligterink2017a,ligterink2018a}. In short, it consists of a central chamber at ultra-high vacuum conditions ($P \leq$ 10$^{-10}$ mbar). A gold-coated, reflective surface is positioned at the center of the chamber, which is cryogenically cooled to temperatures as low as 12~K. Gases are directly deposited (as opposed to background deposition) on this surface, forming an ice layer which simulates the ice mantles on interstellar dust grains. The output of a Microwave Discharge Hydrogen-flow Lamp \citep[MDHL,][and references therein]{ligterink2015} is directed at the surface and used to energetically process the ice with far-UV radiation (10.2 -- 7.3 eV). Chemical changes within the ice are traced by Reflection Absorption IR Spectroscopy (RAIRS) and mass spectrometry in combination with Temperature Programmed Desorption (TPD).

Hydrogen lamps generally have strong emission at the Lyman-$\alpha$ transition at 121.6 nm and H$_{2}$ continuum emission between 140-160 nm. In this work lamp conditions which result in Lyman-$\alpha$ rich or poor emission (see Fig. \ref{fig.VUV_comp}) are used to process the ice samples in order to test the influence of high energy Lyman-$\alpha$ radiation on the solid-state chemistry. In experiments on mixed CH$_{4}$:HNCO ices, far-UV radiation can produce a number of radicals and molecules, such as H, N, HN and CO from HNCO \citep{raunier2004} or CH$_{2}$ and CH$_{3}$ from CH$_{4}$ \citep{bossa2015}. Production of these radicals is influenced by the spectral energy distribution of the lamp. For example, the CH$_{4}$ photo-absorption cross section is high around the Lyman-$\alpha$ transition, but low for wavelengths longer than 140 nm \citep{cruzdiaz2014b}, which will result in limited or no production of CH$_{2}$ and CH$_{3}$ radicals for Lyman-$\alpha$ poor conditions. The total photon flux of the lamp at the position of the ice sample is (1.1$\pm$0.1)\,$\times$10$^{14}$ photons s$^{-1}$ for the Lyman-$\alpha$ rich emission, while it is (6.1$\pm$1.0)\,$\times$10$^{13}$ photons s$^{-1}$ in the Lyman-$\alpha$ poor case.

Gases used during the experiments are CH$_{4}$ (Linde Gas, 99.995\% purity), $^{13}$CH$_{4}$ (Sigma-Aldrich, 99\% purity) and CO (Linde Gas, 99.995\% purity). Regular methane gas contains the natural isotope ratio of $^{12/13}$C of $\sim$90. Throughout this paper regular methane gas will generally be called $^{12}$CH$_{4}$ to emphasize the mass difference in experiments making use of either $^{12}$C or $^{13}$C labelled CH$_{4}$. Gas-phase HNCO is produced by thermal decomposition of cyuranic acid (Sigma-Aldrich, 98\% purity), the solid trimer of HNCO, following a similar protocol as \citet{vanbroekhuizen2004}. Freeze-pump-thaw cycles are used to purify the HNCO sample and mainly remove CO$_{2}$, O$_{2}$ and N$_{2}$ contamination. Hydrogen cyanide (HCN) impurities are sometimes present in the prepared gas, but can not be removed by this technique. Samples of solid acetamide (Sigma-Aldrich, 99\% purity), liquid formamide (Sigma-Aldrich, 99\% purity) and liquid N-methylformamide (Sigma-Aldrich, 99\% purity) are used for verification experiments. 

Gas mixtures are prepared in a gas mixing line by volume mixing, with volumes determined by a gas independent gauge. The mixture is deposited on the substrate at 20~K. Residual gases of the deposition are removed from the chamber during a short waiting period until a pressure of $\sim$1$\times$10$^{-10}$ mbar is reached. Next, the samples are UV irradiated for 20 minutes, corresponding to a total fluence of (1.3$\pm$0.1)\,$\times$10$^{17}$ or (7.3$\pm$1.2)\,$\times$10$^{16}$ photons for the Lyman-$\alpha$ rich and poor case, respectively. Assuming a dark cloud far-UV flux of 10$^{4}$ photons s$^{-1}$ \citep{shen2004}, this matches a dark cloud lifetime of about 3$\times$10$^{5}$ years. After irradiation, Temperature Programmed Desorption (TPD) is employed to linearly heat the sample from 20 to 300~K and let the ice contents desorb from the surface. Material released to the gas-phase is analysed with a high sensitivity Quadrupole Mass Spectrometer (QMS). After deposition, during irradiation and during TPD, IR spectra are recorded at 1 or 2 cm$^{-1}$ resolution using a Fourier Transform InfraRed Spectrometer (FTIRS, 500-4000 cm$^{-1}$), to trace chemical changes in the ice. An overview of the performed irradiation experiments is given in Table \ref{tab.exper}. A number of experiments makes use of a CO matrix that isolates produced radicals and can act as a medium to thermalise hot, energetic, H-atoms. 

\subsection{RAIR spectroscopy analysis method}

RAIR spectra of the experiments are baseline subtracted and IR features are identified by comparing with literature data. Generally small deviations between literature values and this work can arise from differences in transmission versus reflection spectroscopy or matrix effects. The column density ($N_{\rm species}$) of a molecule is determined from the integrated band area ($\int_{\rm band}{\rm log}_{\rm 10}\left(\frac{I_{\rm 0}(\tilde{\nu})}{I(\tilde{\nu})}\right) d\tilde{\nu}$) of an IR feature by:
\begin{equation}
\label{eq.coldens}
N_{\rm species} = \frac{1.1}{3.4}{\rm ln(10)}\frac{\int_{\rm band}{\rm log}_{\rm 10}\left(\frac{I_{\rm 0}(\tilde{\nu})}{I(\tilde{\nu})}\right) d\tilde{\nu}}{A'_{\rm band}},
\end{equation}   
where $A'_{\rm band}$ is the bandstrength of a specific band of a molecule and $\frac{1.1}{3.4}$ is a set-up specific RAIRS scaling factor. Due to longer pathlength through the ice and dipole surface coupling effects, RAIRS has higher sensitivity compared to transmission IR spectroscopy and therefore bandstrength values differ from transmission bandstrength values. For the CryoPAD2 set-up, the bandstrength of the CO stretch mode of carbon monoxide at 2138 cm$^{-1}$ was determined to be 3.4$^{+0.5}_{-0.5} \times 10^{-17}$ cm molecule$^{-1}$ \citep{ligterink2018a}. Using the transmission bandstrength of 1.1$ \times 10^{-17}$ cm molecule$^{-1}$ for the same CO mode \citep{bouilloud2015} and assuming that for identical conditions bandstrengths of other molecules scale approximately in the same way, this scaling factor between RAIRS and transmission IR spectroscopy of $\frac{1.1}{3.4}$ has been derived.

Table~\ref{tab.IRfeat} gives an overview of band positions and used bandstrength values of precursor and expected product species. Most IR parameters are taken from transmission experiments available from literature, with the exception of the CN stretching mode of HCN and the NCO asymmetric stretching mode of CH$_{3}$NCO \citep[][respectively]{gerakines2004,ligterink2017a}. Because no water ice is used in these experiments, the water-poor bandstrength values listed by \citet{vanbroekhuizen2004} are used for HNCO and OCN$^{-}$.

Recently, CH$_{4}$ has been under discussion due to inconsistencies in the literature on the assignment of CH$_{4}$ IR bands to the amorphous or crystalline phase and subsequent deviations in bandstrengths \citep{gerakines2015}. Experiments in this work are conducted with ices at temperatures of 20~K and therefore CH$_{4}$ is considered to be of crystalline nature. Consequently, a crystalline bandstrength value is used, by applying the same correction as performed by \citet{boogert1997} on data recorded by \citet{hudgins1993} to retrieve the bandstrength for the CH$_{4}$ mode at 3010 cm$^{-1}$ as 1.1$\times$10$^{-17}$ cm molecule$^{-1}$. 

\begin{figure*}
\begin{center}
\includegraphics[width=\hsize]{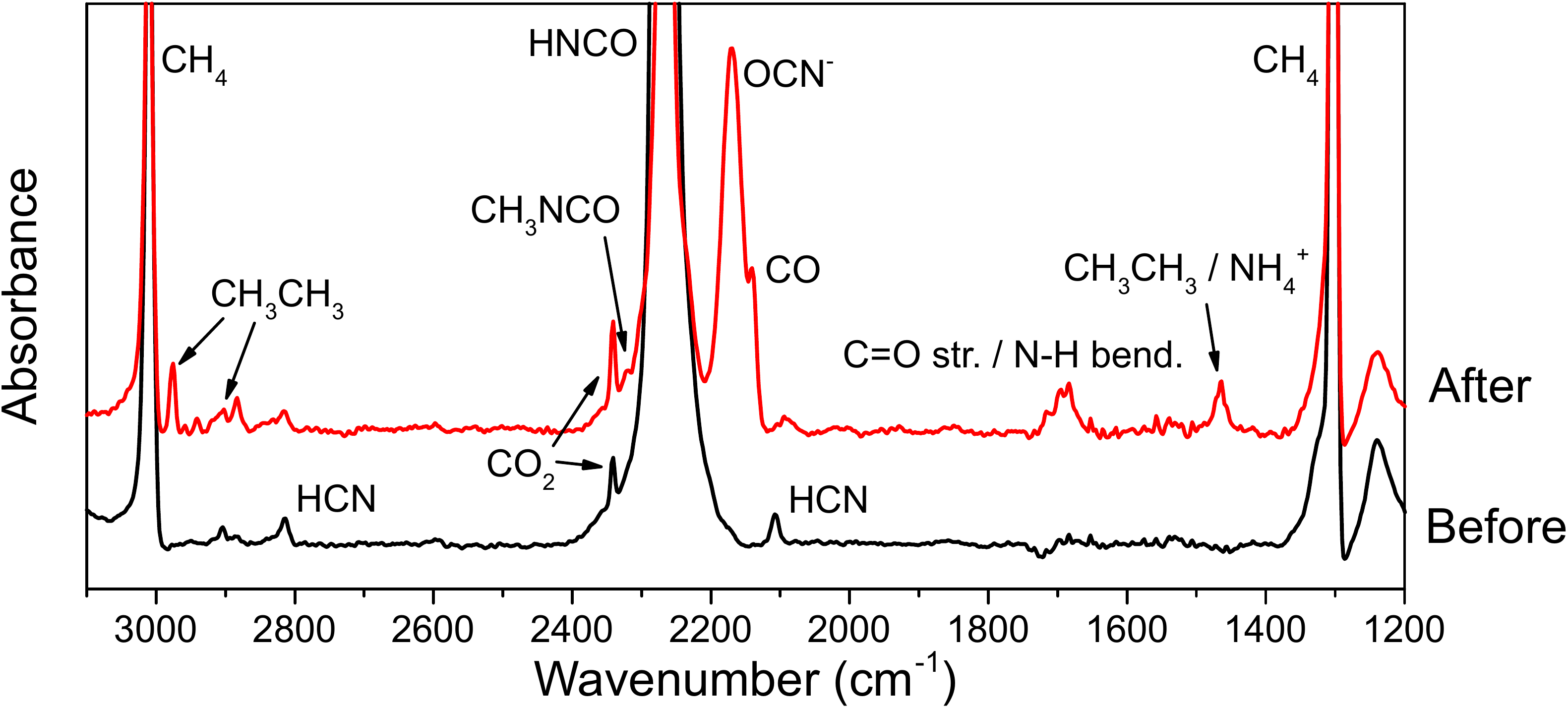}
\caption{IR spectrum between 3100 and 1200 cm$^{-1}$ before (black) and after (red) irradiation of the HNCO:$^{12}$CH$_{4}$ ice mixture (Exp. 1).}
\label{fig.IR_spec}
\end{center}
\end{figure*} 

\subsection{Temperature Programmed Desorption method}

During TPD, precursor and product species desorb from the gold surface and are measured by the QMS, with an ionization source tuned to 70~eV. A temperature ramp of 5 K min$^{-1}$ is used. The mass spectrometric data are corrected for the QMS work function (i.e. the set-up specific response to a certain $m/z$). Products are identified by a combination of their characteristic desorption temperatures and fragmentation patterns. Identification can be hindered if molecules have similar desorption temperatures, fragmentation patterns or co-desorb with other species. Particularly the precursor species HNCO, which desorbs around 130~K, contaminates the TPD signal of other species. Due to its high abundance compared to products, even small fragmentation channels of HNCO, for example at $m/z$ 28 and 29, contribute significantly to product fragmentation patterns and make it impossible to disentangle precursor and product. Therefore, the focus of the TPD data is mainly on the region between 150 and 300~K, avoiding the HNCO desorption peak as much as possible. In addition, isotopic labelling is used to distinguish products by mass shifts. 

In this work the intensity of a fragment at a certain $m/z$ is determined by integrating the area under its baseline subtracted desorption peak. Ratios between two or more molecules can be determined from mass fragments that can be uniquely assigned to a single molecule. Ratios are affected by the fact that products desorbing from the gold substrate can again freeze-out at another cold location, mainly the heat shield ($T$ $\approx$ 70~K). Particularly when comparing the gas-phase ratio of a volatile species like CO ($T_{\rm des}$ = 30~K) with a non-volatile species such as H$_{2}$O ($T_{\rm des}$ = 150~K), the ratio can be offset by freeze-out. Since in this work relatively non-volatile species ($T_{\rm des}$ $\geq$ 100~K) are studied we assume that all these species have a similar freeze-out efficiency and thus the measured ratios between these molecules are not strongly affected. When the fragmentation pattern and electron impact absorption cross section of a molecule are taken into account, the ratio between two molecular mass fragments can be converted in an absolute abundance ratio ($AR$) with the following equation:

\begin{equation}
AR =  \frac{I_{m/z \rm, molecule 1}}{I_{m/z \rm, molecule 2}} \ \frac{\phi_{m/z \rm, molecule 2}}{\phi_{m/z \rm, molecule 1}} \ \frac{\sigma_{\rm molecule 2}}{\sigma_{\rm molecule 1}},
\label{eq.AR}
\end{equation}

\noindent
where $I_{m/z, \rm molecule}$ is the signal intensity of a molecule at a certain fragment $m/z$, $\phi_{m/z, \rm molecule}$ the fragment intensity at this $m/z$ and $\sigma_{\rm molecule}$ the total electron impact absorption cross section of the molecule at 70~eV ionization energy.

\subsection{Fragmentation patterns and desorption temperatures}
\label{sec.fragdes}

Identification of products relies for a large part on molecule specific mass fragmentation patterns and desorption temperatures. Most fragmentation patterns of precursor and possible product species are available in the NIST database\footnote{NIST Mass Spec Data Center, S.E. Stein, director, "Mass Spectra" in NIST Chemistry WebBook, NIST Standard Reference Database Number 69, Eds. P.J. Lindstrom and W.G. Mallard, National Institute of Standards and Technology, Gaithersburg MD, 20899, http://webbook.nist.gov.}, listed for an ionization energy of 70~eV. For HNCO the fragmentation pattern given in \citet{bogan1971} is used. Desorption temperatures of many small species, like CO, water or methanol, have been well studied \citep[e.g.][]{collings2004}. However, for larger species such data are often not available, including the two potential products CH$_{3}$NHCHO and CH$_{3}$C(O)NH$_{2}$. In order to obtain the desorption temperature and fragmentation pattern of these species, samples of pure CH$_{3}$NHCHO and CH$_{3}$C(O)NH$_{2}$ are deposited at 20~K in the CryoPAD2 set-up and their desorption is measured during TPD. Figure~\ref{fig.Tdes} shows their TPD traces at $m/z$ 59, the main detection mass of both molecules. The peak desorption temperature of CH$_{3}$NHCHO is found at 184~K, while CH$_{3}$C(O)NH$_{2}$ desorbs at 219~K. The origin of the smaller peak at $\sim$200~K in the CH$_{3}$C(O)NH$_{2}$ TPD trace is unknown, but could be the result of CH$_{3}$C(O)NH$_{2}$ desorption due to the amorphous to crystalline phase change in the ice. Another important product is NH$_{2}$CHO, for which the desorption temperature is determined to be 210~K (see Fig.~\ref{fig.for_TPD} in Appendix~\ref{ap.TPD_for}). Measured fragmentation patterns of CH$_{3}$NHCHO and CH$_{3}$C(O)NH$_{2}$ are presented in Appendix~\ref{ap.TPD_for}.
     
\section{Results}
\label{sec.res}

The identification of products in the far-UV processed CH$_{4}$:HNCO mixed ices is based on a series of experiments for three primary ice settings. These are a 1:1 HNCO:$^{12}$CH$_{4}$ mixed ice, a pure HNCO ice and a 1:1 HNCO:$^{13}$CH$_{4}$ mixture. Table \ref{tab.exper} lists these as Exps. 1 to 3, together with other experiments performed in this work to test the influence of Lyman-$\alpha$ radiation and CO matrix effects. In the following sections the IR spectroscopic data of Exp. 1 and mass spectrometric data of Exps. 1 to 3 are analysed.

\subsection{IR analysis of UV processed ices}

\begin{figure}
\begin{center}
\includegraphics[width=\hsize]{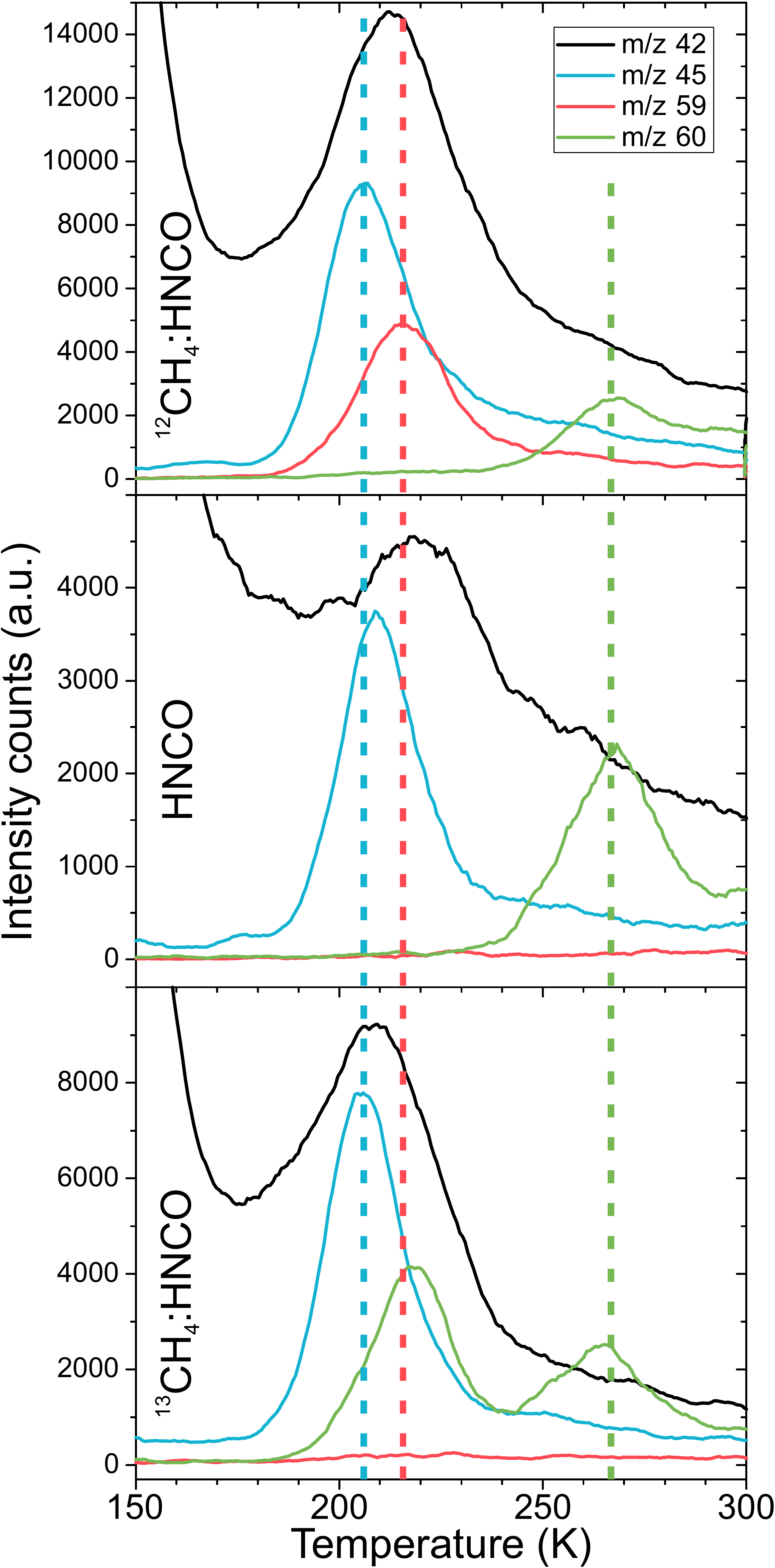}
\caption{TPD traces of $m/z$ 42 (black, NCO fragment of HNCO), 45 (blue, NH$_{2}$CHO), 59 (red, CH$_{3}$C(O)NH$_{2}$) and 60 (green, NH$_{2}$C(O)NH$_{2}$) between 150 and 300~K of far-UV processed HNCO:$^{12}$CH$_{4}$ (Exp. 1, top), pure HNCO (Exp. 2, middle) and HNCO:$^{13}$CH$_{4}$ (Exp. 3, bottom) ice. Dashed lines indicate the desorption peaks of products.}
\label{fig.primary_amides}
\end{center}
\end{figure}

Figure \ref{fig.IR_spec} shows the RAIR spectra between 3100 and 1200 cm$^{-1}$ of the HNCO:CH$_{4}$ mixture (Exp. 1) before and after far-UV irradiation with a total photon fluence of (1.3$\pm$0.1)$\times$10$^{17}$. Before irradiation, bands of HNCO (2266 cm$^{-1}$) and  CH$_{4}$ (3010 and 1302 cm$^{-1}$) are visible. Small features of HCN (2825 and 2108 cm$^{-1}$) and CO$_{2}$ (2341 cm$^{-1}$) are visible as well and are contaminants of the HNCO production process.

After irradiation, bands appear of newly formed products, most of which are found in processed samples of pure CH$_{4}$ or pure HNCO as well. These species are ethane (CH$_{3}$CH$_{3}$), OCN$^{-}$NH$_{4}^{+}$ and CO. The presence of these products indicates that the CH$_{3}$ radical is formed from CH$_{4}$ and that HNCO fragments into CO and NH (or N and H separately). The NH radical reacts sequentially with H atoms to form NH$_{3}$, which engages in an acid-base reaction with HNCO to form the OCN$^{-}$NH$_{4}^{+}$ complex \citep{vanbroekhuizen2004}.

A feature in the wing of the HNCO peak at 2322 cm$^{-1}$ indicates the formation of CH$_{3}$NCO \citep[][]{ligterink2017a,mate2017}. At $\sim$1687 cm$^{-1}$ another feature is visible. In far-UV processing of pure HNCO ice, it was identified as a combination of contributions of the H$_{2}$CO and NH$_{2}$CHO C=O stretch modes and NH bending modes of NH$_{2}$C(O)NH$_{2}$ \citep{raunier2004}. These species likely contribute to this feature, but can not be distinguished. More complex amides, such as CH$_{3}$NHCHO and CH$_{3}$C(O)NH$_{2}$ should also have strong C=O stretch modes that can contribute to this feature. Since other spectroscopic features, characteristic of complex amides, are not seen in the RAIR spectra, mass spectrometry must be used to identify these species. An overview of the identified products visible in the IR in Exp. 1 is given in Table~\ref{tab.IRfeat}. 

\subsection{Identification of the primary amides}

\begin{figure}
\begin{center}
\includegraphics[width=\hsize]{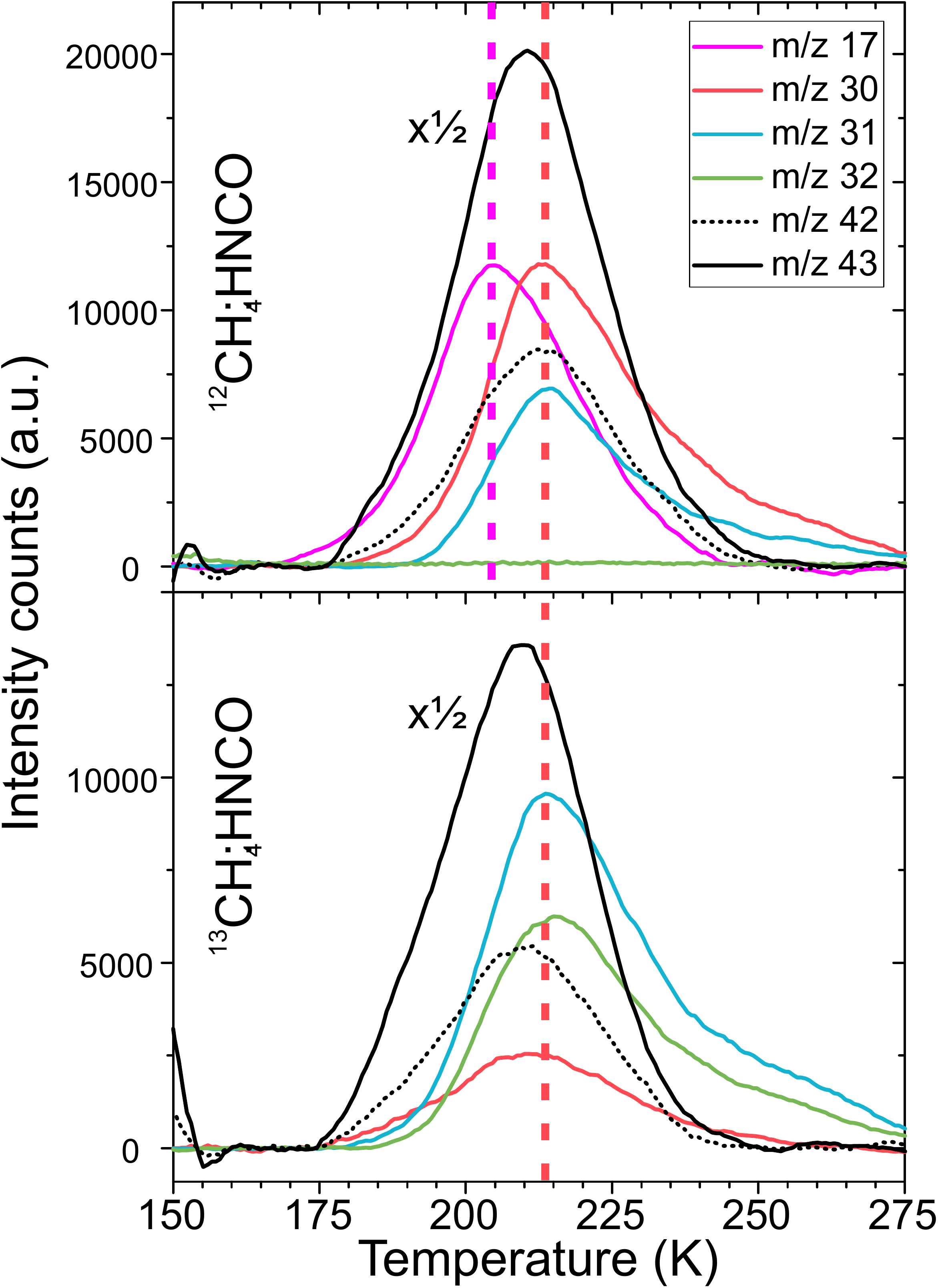}
\caption{Baseline subtracted TPD traces of $m/z$ 17 (purple), 30 (red), 31 (blue), 32 (green), 42 (black dashed) and 43 (black, scaled by factor 1/2) of HNCO:$^{12}$CH$_{4}$ (Exp. 1, top) and HNCO:$^{13}$CH$_{4}$ (Exp. 3, bottom) ice. The dashed purple line indicates the peak desorption temperature of $m/z$ 17 and the red dashed line indicates the peak desorption temperature of $m/z$ 30, 31 and 32.}
\label{fig.methylamine}
\end{center}
\end{figure}

\begin{figure*}
\begin{center}
\includegraphics[width=\hsize]{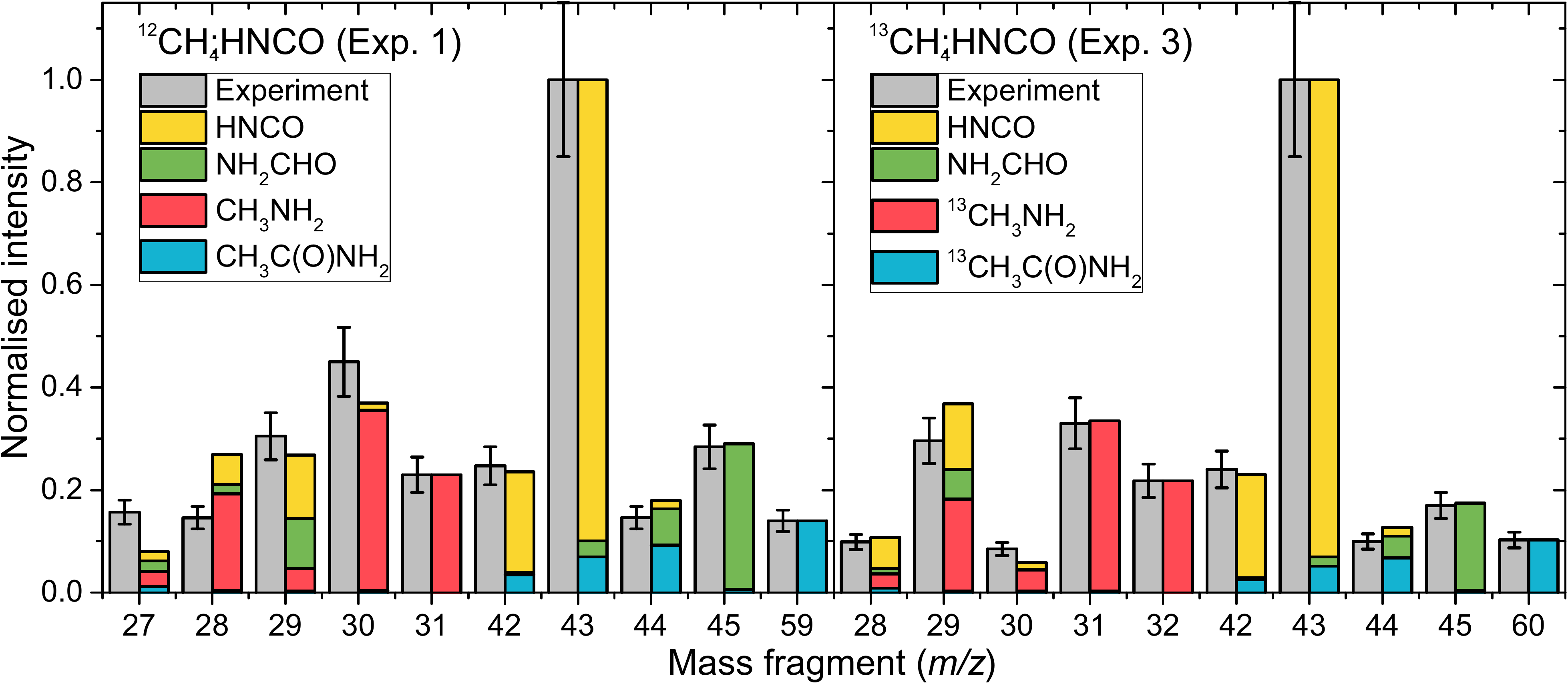}
\caption{The fragments at $m/z$ 60/59, 45, 44, 43, 42, 32/31, 31/30, 30/29, 29/28 and 28/27 found desorbing around 210~K in experiments 1 and 3 (black), fitted with the combined fragmentation patterns of $^{12/13}$acetamide (blue), $^{12/13}$methylamine (red), formamide (green) and HNCO (yellow). Note that intensity ratios do not directly reflect abundance ratios.}
\label{fig.meth_fit}
\end{center}
\end{figure*}

Figure \ref{fig.primary_amides} shows the TPD traces between 150 and 300~K of the main masses of the simplest, or primary, amides that can be formed from HNCO:CH$_{4}$ ice mixtures. These masses are $m/z$ 45 for NH$_{2}$CHO, $m/z$ 59 for either CH$_{3}$C(O)NH$_{2}$ or CH$_{3}$NHCHO and $m/z$ 60 for NH$_{2}$CONH$_{2}$. The secondary mass channel of HNCO, $m/z$ 42, is included as well to trace HNCO. The panels show from top to bottom the results of the UV irradiation of $^{12}$CH$_{4}$:HNCO, HNCO and $^{13}$CH$_{4}$:HNCO (Exps. 1--3, respectively). 

In each of the panels a prominent trailing slope of $m/z$ 42 between 150 and 300~K, with a superimposed desorption feature at $\sim$210~K is seen. The trailing slope is due to residual gas of the main HNCO desorption peak at 130~K, see Fig. \ref{fig.full} in Appendix \ref{ap.TPD_for}. The desorption feature is caused by the thermal decomposition of the OCN$^{-}$NH$_{4}^{+}$ (or another cation) salt complex and subsequent desorption of HNCO.

Three desorption peaks of $m/z$ 45, 59 and 60 are visible at $\sim$205, $\sim$215 and $\sim$265~K, respectively. Moreover, $m/z$ 45 and 60 show up in all panels, including the irradiated pure HNCO ice, and therefore must be photoproducts directly resulting from HNCO. The desorption peak of $m/z$ 45 at 205~K matches the desorption temperature of pure NH$_{2}$CHO at 210~K and is therefore assigned to this molecule. The position of the $m/z$ 60 desorption peak is consistent with TPD traces of NH$_{2}$C(O)NH$_{2}$ obtained by \citet{forstel2016} and is identified accordingly. Both identifications are consistent with results of pure HNCO irradiation by \citet{raunier2004}. 

The $m/z$ 59 feature is the result of a reaction between CH$_{4}$ and HNCO related fragments, as can be inferred from its non-presence in the pure HNCO experiment and 1 amu mass shift to $m/z$ 60 in the $^{13}$CH$_{4}$:HNCO experiment. This product can either be CH$_{3}$NHCHO or CH$_{3}$C(O)NH$_{2}$. Other isomers like acetaldoxime (CH$_{3}$CHNOH) and nitrosoethane (CH$_{3}$CH$_{2}$NO) are deemed unlikely to be responsible for the $m/z$ 59 feature, due to the many fragmentation and reaction steps that need to be invoked to form these products. The desorption peak of $m/z$ 59 at 215~K is close to the desorption temperature of pure CH$_{3}$C(O)NH$_{2}$ at 219~K, as shown in Fig. \ref{fig.Tdes}. Trapping of CH$_{3}$NHCHO in the ice contents (mainly OCN$^{-}$NH$_{4}^{+}$ and NH$_{2}$CHO) and thus shifting to a higher desorption temperature is ruled out, due to the relatively volatile nature of CH$_{3}$NHCHO. If this were the case, one would expect it to desorb before or with NH$_{2}$CHO and OCN$^{-}$NH$_{4}^{+}$, but a desorption at a higher temperature, as is the case here, is unlikely.

\subsection{Mass fragmentation pattern fit}

Further evidence for the identification of CH$_{3}$C(O)NH$_{2}$ can be found by fitting its mass fragmentation pattern. As shown in section~\ref{sec.fragdes} and Fig.~\ref{fig.NISTcomp}, CH$_{3}$C(O)NH$_{2}$ has prominent fragmentation channels at $m/z$ 44 and 43, while CH$_{3}$NHCHO has channels into $m/z$ 30 and 29. Interestingly, $m/z$ 30 does show up prominently in the TPD trace around 220~K, as does $m/z$ 31, see Fig.~\ref{fig.methylamine}. Both these masses are shifted to $m/z$ 31 and 32, respectively, in the $^{13}$CH$_{4}$ isotope experiment. The $m/z$ 30 and 31 signals are unlikely to be from CH$_{3}$NHCHO, however. First, there is no fragment channel at $m/z$ 31 for CH$_{3}$NHCHO (although $m/z$ 30 and 31 do not necessarily have to be associated). Second, the $m/z$ 30/59 ratio in the experiments is greater than 1, while the $m/z$ 30/59 ratio of pure CH$_{3}$NHCHO is 0.21. Therefore, these masses are most likely due to another species, presumably methylamine (CH$_{3}$NH$_{2}$). 

To strengthen this claim, the main masses desorbing around 200-220~K are fitted with the fragmentation patterns of HNCO, NH$_{2}$CHO, CH$_{3}$NH$_{2}$ and CH$_{3}$C(O)NH$_{2}$ (see Table~\ref{tab.frag_comp} in Appendix~\ref{ap.TPD_for} for the fragmenation patterns of these molecules). These masses are integrated between 170 and 250~K, and in sequence the four components are added to reproduce the experimental fragment patterns. Figure~\ref{fig.meth_fit} shows the results of these fits for Exps. 1 and 3. The ratios between the fitted components CH$_{3}$NH$_{2}$:HNCO:NH$_{2}$CHO:CH$_{3}$C(O)NH$_{2}$ are 28:43:17:12 for the $^{12}$C experiment and 30:50:11:9 for the $^{13}$C experiment.

The experimental mass patterns are reasonably well fitted with a combination of these four molecules, particularly in the isotope experiment. Small differences can still occur due to contributions of other species to $m/z$ 27, 28, 29 and 30, such as fragments originating from more complex molecules. This fit confirms the identification of CH$_{3}$C(O)NH$_{2}$ and makes it very likely that CH$_{3}$NH$_{2}$ is also formed in these experiments. In pure form CH$_{3}$NH$_{2}$ thermally desorbs at relatively low temperatures of 100-120~K \citep[e.g.][]{chaabouni2018}, while in these experiments it desorbs at $\sim$220~K. Due to bulk ice release with HNCO it is unclear in the TPD traces if some CH$_{3}$NH$_{2}$ releases around 110~K. The release of CH$_{3}$NH$_{2}$ around 220~K can be explained by the formation of an OCN$^{-}$CH$_{3}$NH$_{3}^{+}$ salt complex, which has a higher desorption temperature, similar to the OCN$^{-}$NH$_{4}^{+}$ complex. Reference data on the thermal decomposition of the OCN$^{-}$CH$_{3}$NH$_{3}^{+}$ complex is needed to confirm this. Different peak desorption temperatures between these two salts are explained by the fact that CH$_{3}$NH$_{2}$ is a stronger base than NH$_{3}$ and therefore more thermal energy is required to dissociate this complex.

\subsection{Secondary amides and larger species}

\begin{figure}
\begin{center}
\includegraphics[width=\hsize]{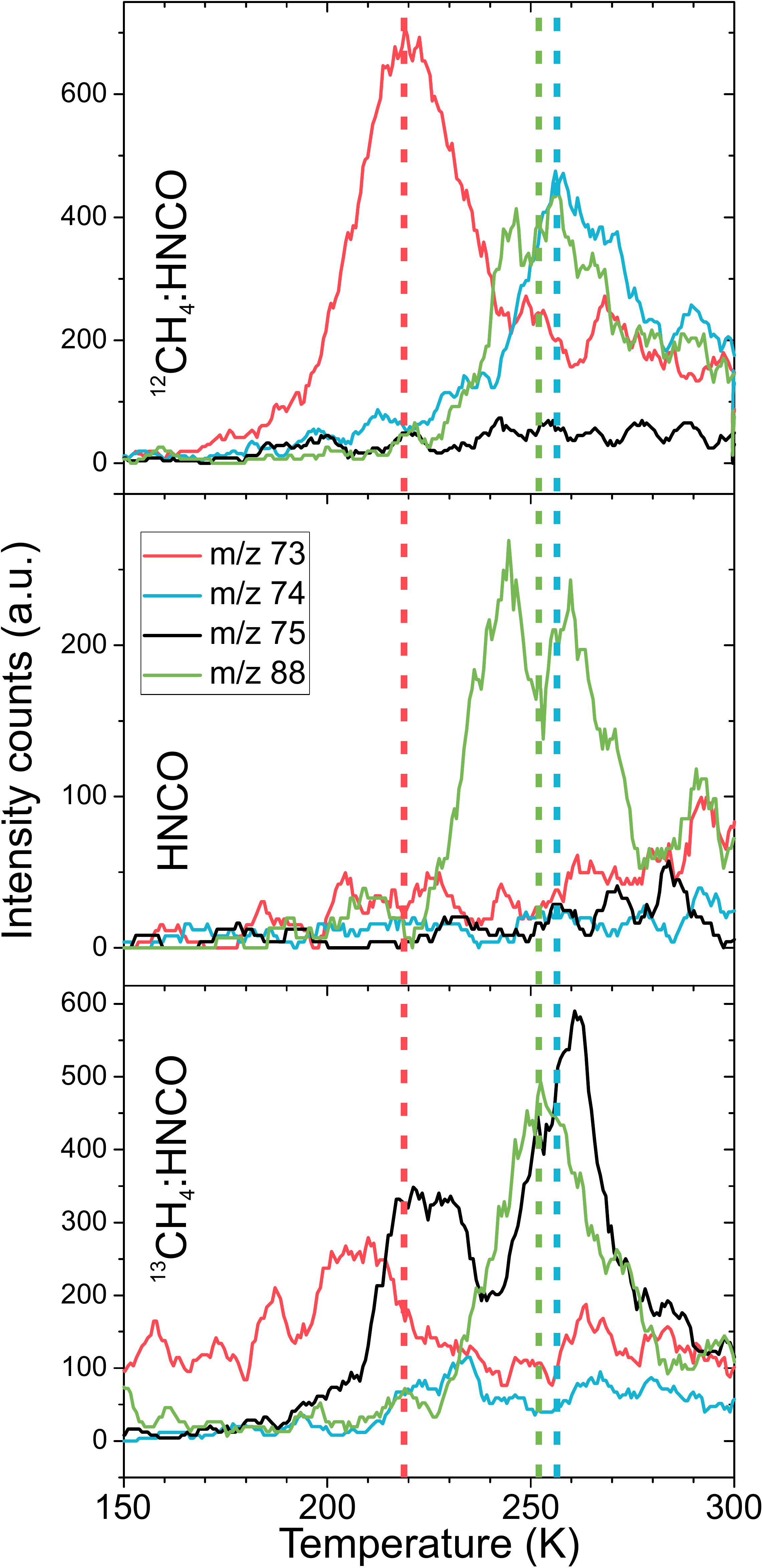}
\caption{TPD of $m/z$ 73 (red), 74 (blue), 75 (black) and 88 (green) between 150 and 300~K of UV processed HNCO:$^{12}$CH$_{4}$ (top), HNCO (middle) and HNCO:$^{13}$CH$_{4}$ (bottom) ices. Dashed lines indicate the desorption peaks of various products.}
\label{fig.highmass}
\end{center}
\end{figure}

Evidence for the formation of more complex species is seen, as shown in Fig.~\ref{fig.highmass}. Three features at $m/z$ 73, 74 and 88 are detected in the TPD traces of Exps. 1--3. The first feature of $m/z$ 73, desorbing at $\sim$220~K, does not show up in the processed pure HNCO sample. In the $^{13}$CH$_{4}$ experiment the feature shifts to $m/z$ 75, indicating that two $^{13}$CH$_{x}$ groups are part of this product. The $m/z$ 74 feature, desorbing at $\sim$260~K, also does not show up in the pure HNCO processed ice and shifts by one mass to $m/z$ 75 in the $^{13}$CH$_{4}$ mixture. Therefore this product incorporates only one $^{13}$CH$_{x}$ group. $m/z$ 88 is seen in all three the panels and desorbs around 250~K. Since it is observed in the pure HNCO ice and no isotope shifts are seen, it likely results from two HNCO related intermediates. 

We note that the low signal-to-noise ratio (compare to Figs. \ref{fig.primary_amides} and \ref{fig.methylamine}) limits the identification of molecules and therefore contributions of more than three species to these $m/z$ signals can not be ruled out. Also, desorption temperatures of candidate molecules are not available and matching fragmentation patterns is hindered due to the low signal and overlap in fragmentation channels of previously identified, more abundant, species. Nevertheless, some candidates can be suggested, especially when assuming these high mass species are derived from, or related to the first generation of amides. For the $m/z$ 73 signal propionamide (CH$_{3}$CH$_{2}$C(O)NH$_{2}$), N-methylacetamide (CH$_{3}$NHCOCH$_{3}$) and dimethylformamide ((CH$_3$)$_{2}$NCHO) could be responsible. The latter two seem unlikely because CH$_{3}$NHCHO is not detected in these mixtures. The fact that CH$_{3}$CH$_{3}$ is identified in the IR spectra of these experiments, favors the assignment as CH$_{3}$CH$_{2}$C(O)NH$_{2}$. The two most likely options for $m/z$ 74 are methylcarbamide (CH$_{3}$NHC(O)NH$_{2}$) and 2-amino acetamide (NH$_{2}$CH$_{2}$C(O)NH$_{2}$). Again, based on the non-detection of CH$_{3}$NHCHO, the latter molecule is the more likely candidate. Finally, for $m/z$ = 88 two options exist; oxamide (NH$_{2}$-C(O)-C(O)-NH$_{2}$) or 1,2-hydrazinedicarboxaldehyde (CHO-NH-NH-CHO). 

\subsection{Comparison of experimental conditions}

\begin{figure*}
\begin{center}
\includegraphics[width=\hsize]{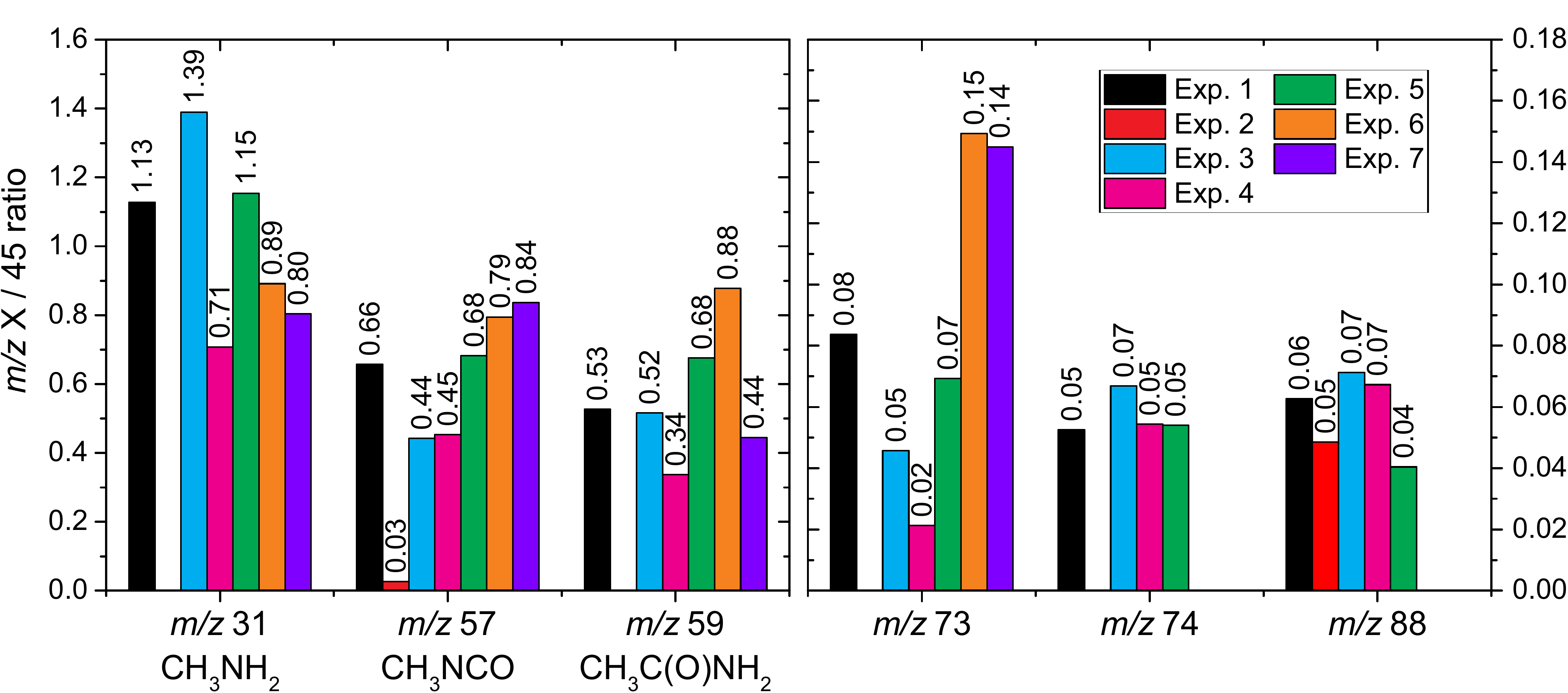}
\caption{Comparison of mass fragment intensity ratios in all experiments. Note that intensity ratios do not reflect molecular abundance ratios.}
\label{fig.comp}
\end{center}
\end{figure*}

Besides the experiments analysed thus far (Exps. 1--3), the influence of Lyman-$\alpha$ rich or poor emission (Exps. 1--4 \& 5--7) and the presence of a CO matrix (Exps. 6--7) on the chemistry has been investigated, see Table~\ref{tab.exper}. In this section trends and variations between all experiments (Exps. 1--7) are investigated based on a number of mass fragments, see Fig.~\ref{fig.comp}. In this figure, ratios of prominent mass fragments are given with respect to $m/z$ 45 (NH$_{2}$CHO). Here, $m/z$ 57 (CH$_{3}$NCO) is included, but note that this species co-desorbs below 150~K, with HNCO. The experiments with the CO matrix (Exps. 6 and 7) are less reliable in tracing $m/z$ 31, because not desorption peaks, but rather desorption plateaus are seen for this mass. The same effect is observed for $m/z$ 30. Finally, results of the $^{13}$CH$_{4}$ experiment (Exp. 3) are included at the non-isotope masses presented in this figure. 

\begin{figure*}
\begin{center}
\includegraphics[width=\hsize]{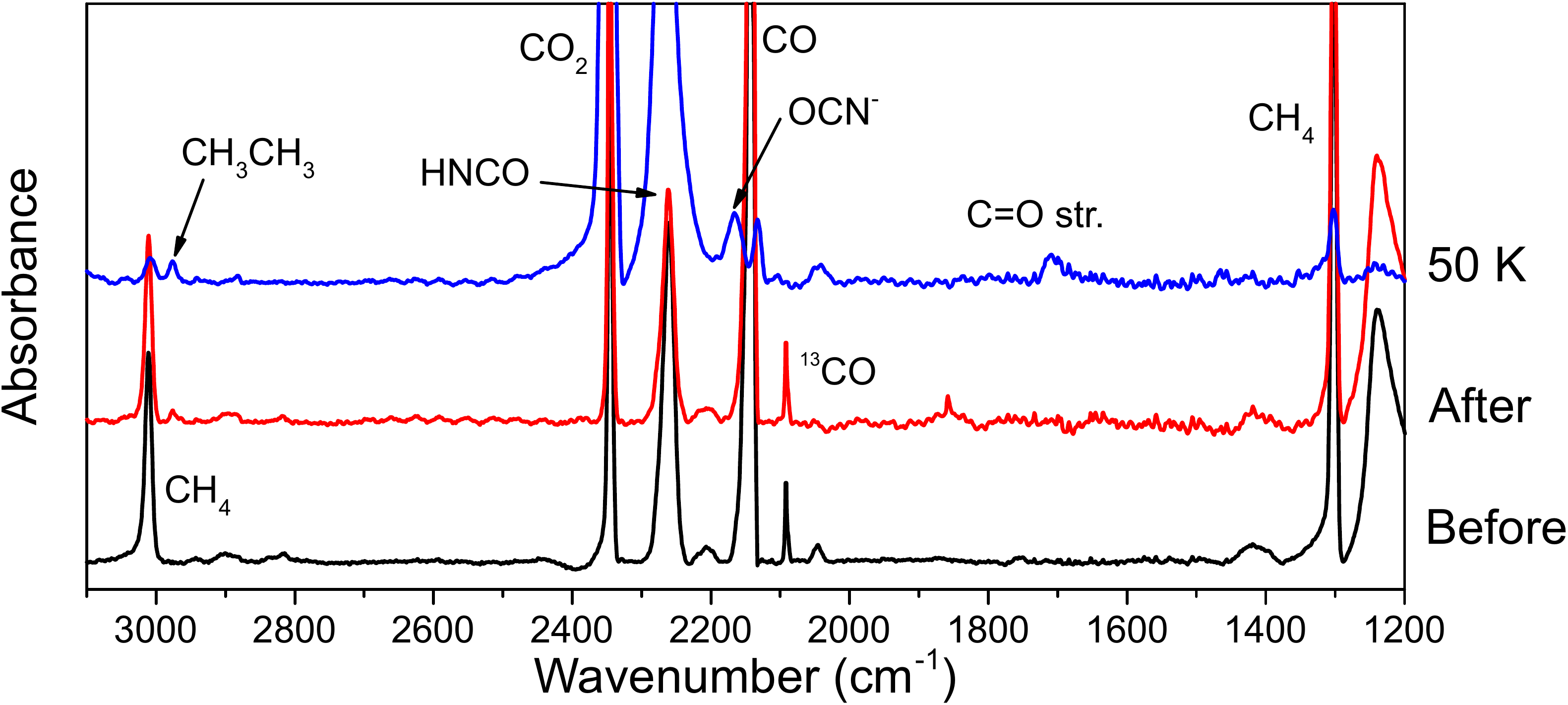}
\caption{Spectra of the CH$_{4}$:HNCO:CO experiment (Exp. 7), before irradiation (black), after irradiation (red) and after heating to 50~K (blue).}
\label{fig.IR_spec_therm}
\end{center}
\end{figure*}

From Fig.~\ref{fig.comp} it becomes clear that CH$_{3}$NH$_{2}$, CH$_{3}$NCO and CH$_{3}$C(O)NH$_{2}$ are present in all experiments, masses $m/z$ 73, 74 and 88 in most and intensity ratios are relatively similar. The presence of Lyman-$\alpha$ (Exps. 5--7) is not a prerequisite for the formation of these species. The CO matrix (Exps. 6 and 7) does hinder the formation of the products associated with $m/z$ 74 and 88, however. Since CH$_{4}$ mainly absorbs far-UV radiation around the Lyman-$\alpha$ transition \citep{cruzdiaz2014b}, it either implies that the small amount of available high-energy radiation is enough to make a sufficient amount of CH$_{3}$ radicals, or this radical is formed efficiently by H-abstraction reactions with other atoms and molecules, such as N and NH \citep[see also discussion in][]{bossa2015}. The experiments conducted in a CO matrix hint that not hydrogenation, but radical recombination reactions are responsible for forming the products. The H-atoms produced in the photodestruction of the precursor species are thermalised by the interaction with the CO matrix and thus have no, or limited, excess energy. Since hydrogenation with non-energetic H-atoms has been shown not to be able to hydrogenate HNCO into NH$_{2}$CHO \citep{noble2015} at low (10--17~K) temperatures, hydrogenation of HNCO can be ruled out in these experiments. Because roughly similar product ratios are found in experiments with and without a CO matrix, energetic H-atom addition to HNCO is unlikely to be the mechanism to form NH$_{2}$CHO and intermediate radicals. Also, IR spectra show no formation of products during irradiation, but do show a decrease in the HNCO and CH$_{4}$ column. Upon warm-up of the ice, formation of products is seen, specifically above 30~K when CO desorbs, see Fig.~\ref{fig.IR_spec_therm}.

\section{Interstellar observations}
\label{sec.obs}

Detections of CH$_{3}$C(O)NH$_{2}$ and a tentative identification of CH$_{3}$NHCHO have been made towards high-mass Young Stellar Objects (YSOs), specifically Sgr B2 and Orion KL \citep{hollis2006,halfen2011,cernicharo2016,belloche2017}. Detections towards low-mass protostars are lacking so far, and this may be due to physical differences between high- and low-mass sources caused by, for example, dust grain temperatures and radiation fields. IRAS~16293--2422 (hereafter IRAS~16293) is such a low-mass protostellar source, consisting of two protostars A and B. Due to its relative close proximity at 120 pc and high luminosity of 21 $L_{\rm \odot}$ \citep{jorgensen2016} this has been a long time favourite low-mass object to study complex chemistry \citep[e.g.][]{cazaux2003}. For this reason, we have searched for CH$_{3}$C(O)NH$_{2}$ and CH$_{3}$NHCHO towards the protostar IRAS~16293B.

\subsection{Observations and analysis}

\begin{figure*}
\begin{center}
\includegraphics[width=\hsize]{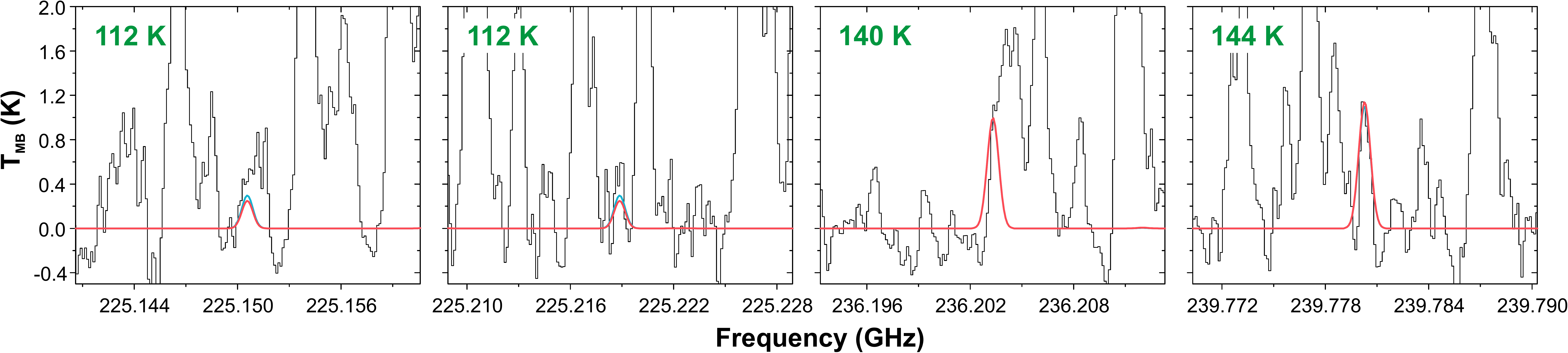}
\caption{Spectrum of IRAS~16293--2422B (black) with the synthetic spectrum of CH$_{3}$C(O)NH$_{2}$ at $T_{\rm ex}$ = 100~K (blue) and 300~K (red) overplotted. On the y-axis, the intensity of the measured rotational lines is indicated by the main beam temperature ($T_{\rm MB}$) in Kelvin. The upper state energy of each line is given in green.}
\label{fig.acetamide_line}
\end{center}
\end{figure*}

Interferometric observations of the Atacama Large Millimeter/submillimeter Array (ALMA) are used for this search. Specifically, data from the Protostellar Interferometric Line Survey \citep[PILS,][]{jorgensen2016}, supplemented by data from Taquet et al. (subm., project-id: 2016.1.01150.S). are used. In short, these data cover large parts of the spectral ranges in ALMA Bands 6 and 7 at a spectral resolution of 0.2~km~s$^{-1}$ and high rms sensitivity of 0.5--5 mJy beam$^{-1}$ km s$^{-1}$. The observations have a circular restoring beam of 0$.\!\!^{\prime\prime}$5, which ensures that the hot corino around IRAS~16293B (diameter of $\sim$160 AU) can be spatially resolved. A position located at 1 beam size offset with respect to the peak dust continuum emission of source B is used for the analysis. This region is characterised by narrow rotational lines of $\sim$1~km~s$^{-1}$ and therefore line-blending is minimized \citep[e.g.][]{coutens2016,lykke2017}. 

The rotational spectroscopic data of the lowest three vibrational states of CH$_{3}$C(O)NH$_{2}$ are provided by \cite{ilyushin2004} and the rotational spectroscopy of CH$_{3}$NHCHO is given by \citet{belloche2017}, respectively. The Jet Propulsion Laboratory (JPL\footnote{\url{http://spec.jpl.nasa.gov}}) catalog for molecular spectroscopy \citep{Pickett1998} and Cologne Database for Molecular Spectroscopy \citep[CDMS;][]{muller2001,muller2005} are used to check for contaminating lines. The CASSIS line analysis software\footnote{\url{http://cassis.irap.omp.eu/}} is used to analyse the spectra.

Upon detection of rotational lines belonging to these molecules, their column density is derived by making a grid of synthetic spectra, assuming Local Thermodynamic Equilibrium (LTE), and determining the best fit model based on the minimum $\chi^{2}$. The column density ($N$) is scanned between 5$\times$10$^{14}$ and 5$\times$10$^{15}$ cm$^{-2}$ in steps of 1$\times$10$^{14}$ cm$^{-2}$, while $T_{\rm ex}$ is fixed at either 100 or 300~K, the excitation temperature at which most complex molecules are found in the PILS data. Other input parameters for the synthetic spectra are kept constant, such as the line width at 1~km~s$^{-1}$ and $V_{\rm lsr}$ at 2.7~km~s$^{-1}$.

\subsection{Results}

In the observed spectra, two unblended and two partially blended lines are identified belonging to CH$_{3}$C(O)NH$_{2}$, see Fig. \ref{fig.acetamide_line}, making for a tentative identification of the molecule. The parameters of the detected lines, and molecules they are blended with, are presented in Table~\ref{tab.transitions} in Appendix~\ref{ap.spec}. The lines are low in intensity and therefore quite sensitive to baseline fluctuations. At the same time, their upper state energies only cover a small region of $E_{\rm up}$ = 112 -- 141~K. These issues make it difficult to properly fit a synthetic spectrum and derive an excitation temperature. We therefore run our grid of synthetic spectra at fixed temperatures of 100 and 300~K, similar to \citet{coutens2016} and \citet{ligterink2017a}. The best-fit column densities are found to be 9$\times$10$^{14}$ and 2.5$\times$10$^{15}$ cm$^{-2}$ for $T_{\rm ex}$ = 100 and 300~K, respectively (Fig. \ref{fig.acetamide_line}). 

No CH$_{3}$NHCHO lines are identified and therefore an upper limit column density is provided. From line-free regions where CH$_{3}$NHCHO lines are expected, the column density is determined to be $\leq$1$\times$10$^{15}$ cm$^{-2}$ at $T_{\rm ex}$ of 100--300~K.

The comparison between both species gives [CH$_{3}$NHCHO] / [CH$_{3}$C(O)NH$_{2}$] $\leq$0.4--1.1 in IRAS~16293B, consistent with the ratio of 0.7 derived towards Sgr B2(N2) \citep{belloche2017}. For both species column density ratios are determined with respect to relevant molecules. Comparisons are made with HNCO (3$\times$10$^{16}$ cm$^{-2}$), NH$_{2}$CHO \citep[1$\times$10$^{16}$ cm$^{-2}$,][]{coutens2016}, CH$_{3}$NCO \citep[3-4$\times$10$^{15}$ cm$^{-2}$,][]{ligterink2017a} and H$_{2}$ \citep[$\geq$1.2$\times$10$^{25}$ cm$^{-2}$,][]{jorgensen2016}. The column densities of the first four molecules have been determined towards the 1 beam offset position towards source B. The H$_{2}$ column density is determined from dust continuum emission \citep{jorgensen2016}, which is optically thick at the 1 beam offset position and thus a lower limit. Abundances for the molecules are given in Table~\ref{tab.obs}. 

\begin{table}
     \caption[]{Observed CH$_{3}$C(O)NH$_{2}$ and CH$_{3}$NHCHO abundances with respect to HNCO, NH$_{2}$CHO, CH$_{3}$NCO, and H$_{2}$ towards IRAS 16293--2422B.}
         $$
         \begin{tabular}{l l l}
            \hline
            \hline
            \noalign{\smallskip}
            Reference & CH$_{3}$C(O)NH$_{3}$ & CH$_{3}$NHCHO \\
            \noalign{\smallskip}
            \hline
            \noalign{\smallskip}
			/HNCO 			& $\sim$0.03--0.08	& $\leq$0.03 \\			
			/NH$_{2}$CHO	& $\sim$0.09--0.25	& $\leq$0.10 \\
			/CH$_{3}$NCO	& $\sim$0.23--0.83	& $\leq$0.25 \\
			/H$_{2}$		& $\leq$2.10$\times$10$^{-10}$	& $\leq$8.3$\times$10$^{-11}$ \\
            \noalign{\smallskip}
            \hline
            
         \end{tabular}     
         $$     
	\label{tab.obs}
\end{table}

\subsection{Experimental and observational abundance comparison}

\begin{figure}
\begin{center}
\includegraphics[width=\hsize]{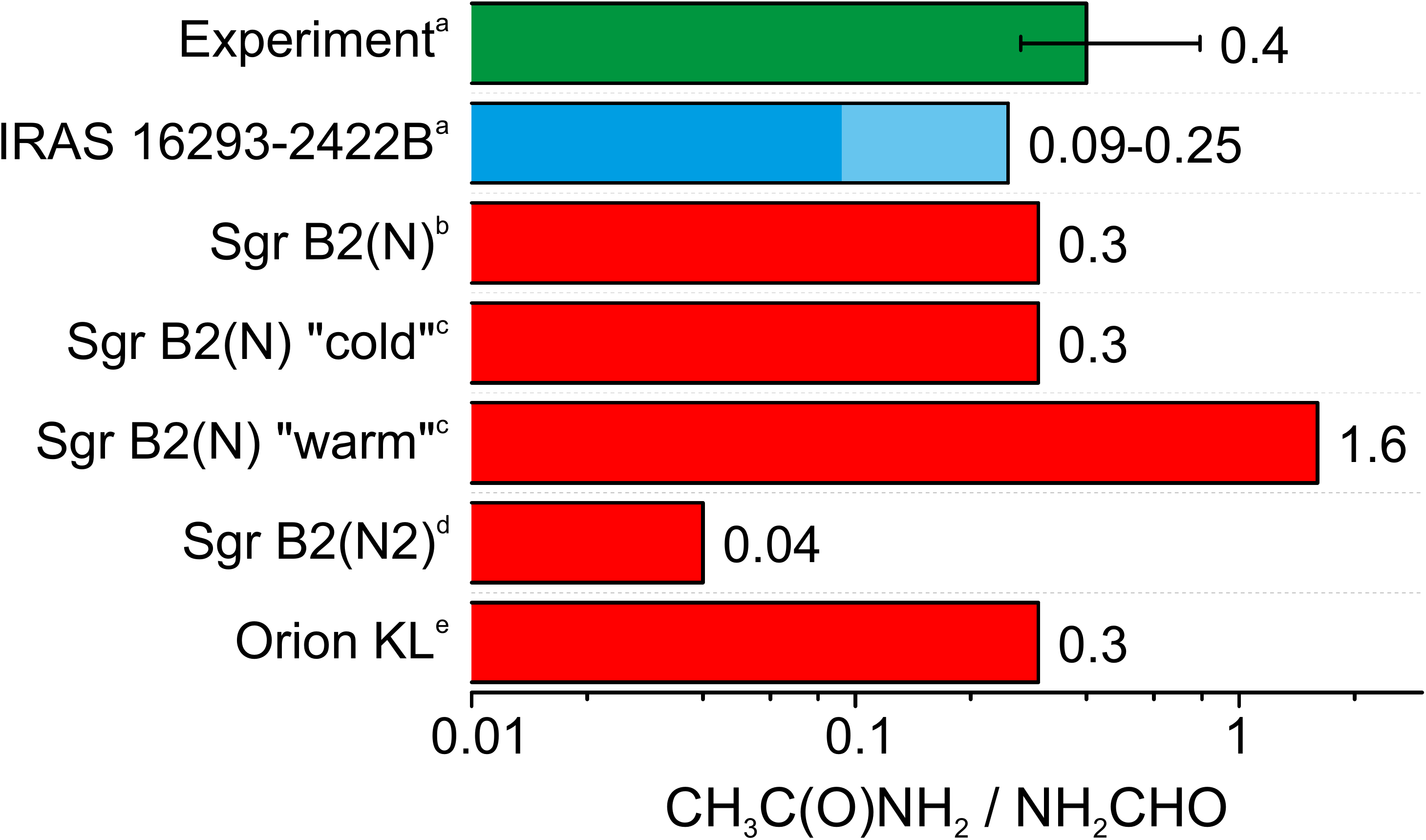}
\caption{CH$_{3}$C(O)NH$_{2}$ over NH$_{2}$CHO abundance ratio as found in Exp. 1 of this work compared with ratios derived in observational studies. $^{a}$This work; $^{b}$\citet{hollis2006}; $^{c}$\citet{halfen2011}; $^{d}$\citet{belloche2017}; $^{e}$\citet{cernicharo2016}.}
\label{fig.AR_acetamide}
\end{center}
\end{figure}

TPD data from the experiments yield ratios between molecules that can be directly compared to observed gas-phase ratios and make it possible to directly link laboratory with interstellar chemical processes. Strictly speaking, experiments and observations cannot be compared directly due to differences in, for example, irradiation fluxes, timescales and exact ice composition. Nevertheless, the experiments do provide insight in abundance ratio trends.Equation \ref{eq.AR} is used to determine the experimental gas-phase ratios for CH$_{3}$C(O)NH$_{2}$ and CH$_{3}$NH$_{2}$ relative to NH$_{2}$CHO. The $I_{\rm 59, acetamide}$/$I_{\rm 45, formamide}$ ratio is taken as 0.50, while that of $I_{\rm 31, methylamine}$/$I_{\rm 45, formamide}$ is 1.10. Fragmentation patterns give $\phi_{\rm 59, acetamide}$ = 0.29, $\phi_{\rm 45, formamide}$ = 0.41 and $\phi_{\rm 31, methylamine}$ = 0.26. To our knowledge, total electron impact absorption cross sections have not experimentally been determined for any of these species. The theoretical absorption cross section for NH$_{2}$CHO is calculated to be 5.595 \r{A}$^{2}$ \citep{gupta2013}, but calculated values for CH$_{3}$C(O)NH$_{2}$ and CH$_{3}$NH$_{2}$ are not available. Therefore we adopt the calculated cross sections of CH$_{3}$NHCHO (10.063 \r{A}$^{2}$) and propylamine (CH$_{3}$CH$_{2}$CH$_{2}$NH$_{2}$, 7.569 \r{A}$^{2}$) from \citet{gupta2013} with a generous error bar of $\pm$5 \r{A}$^{2}$. Based on these numbers, ratios are calculated to be [CH$_{3}$C(O)NH$_{2}$] / [NH$_{2}$CHO] = 0.4$^{+0.39}_{-0.14}$ and [CH$_{3}$NH$_{2}$] / [NH$_{2}$CHO] = 1.32$^{+2.56}_{-0.53}$.

\begin{figure}
\begin{center}
\includegraphics[width=\hsize]{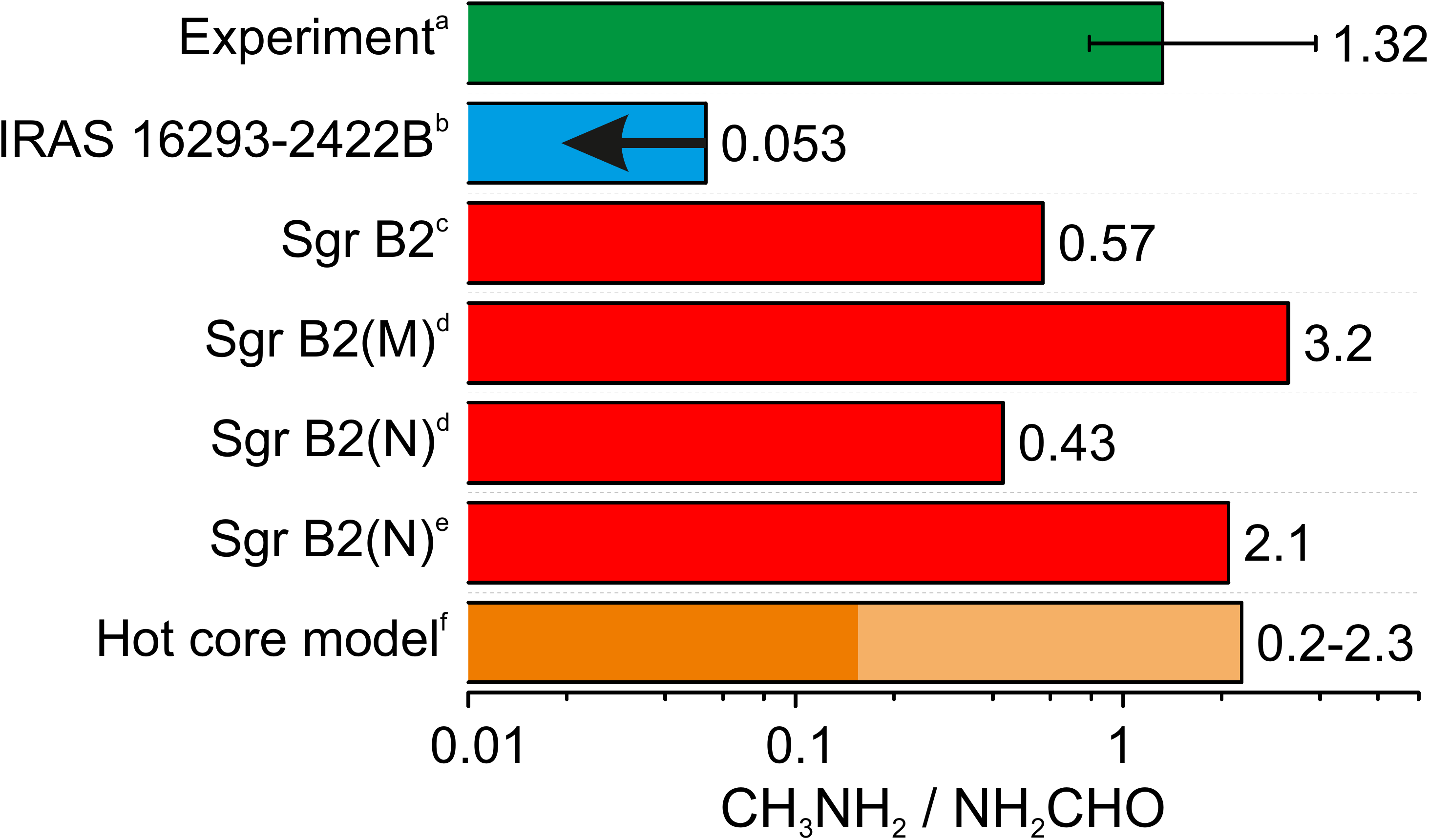}
\caption{CH$_{3}$NH$_{2}$ over NH$_{2}$CHO abundance ratio as found in Exp. 1 of this work compared with ratios derived in observational studies and calculated in models. $^{a}$This work; $^{b}$Ligterink et al. 2018 subm.; $^{c}$\citet{turner1991}; $^{d}$\citet{belloche2013}; $^{e}$\citet{neill2014}; $^{f}$\citet{garrod2013}.}
\label{fig.AR_methylamine}
\end{center}
\end{figure}

The experimentally derived CH$_{3}$C(O)NH$_{2}$ abundance is compared with observational ratios towards Sgr B2, Orion KL and IRAS~16293B in Fig.~\ref{fig.AR_acetamide}. The observational data are a mix of single dish \citep{hollis2006,halfen2011} and interferometric ALMA \citep[this work,][]{cernicharo2016,belloche2017} data. When comparing with single dish data, one needs to be aware that beam dilution effects can strongly affect column densities and thus molecule ratios. In general the [CH$_{3}$C(O)NH$_{2}$] / [NH$_{2}$CHO] ratios are found to be similar within a factor of a few, including the tentative detection towards IRAS~16293B. A clear outlier is the ratio of 0.04 derived for Sgr B2(N2) \citep{belloche2017}. The overall match between observational and experimental data hints that similar ice processes on interstellar dust grains and in laboratory ices are responsible for forming CH$_{3}$C(O)NH$_{2}$ and NH$_{2}$CHO and thus both species have a common chemical origin. Furthermore, the mechanism that is at play is not significantly affected by the colder and radiation poorer conditions of IRAS~16293B.

In Fig.~\ref{fig.AR_methylamine}, the [CH$_{3}$NH$_{2}$] / [NH$_{2}$CHO] ratios are shown for the experimental results, observational data towards Sgr~B2 \citep{turner1991,belloche2013,neill2014} and IRAS~16293B \citep{ligterink2018b} and various hot core models by \citet{garrod2013}. In general, a ratio of $\sim$1 is found, except for the abundance towards the low-mass protostar, which has at least an order of magnitude lower abundance. \citet{ligterink2018b} indicates that the difference in low- and high-mass hot cores may be caused by differences in physical conditions, such as grain temperatures and intensity of the radiation field. The experimental conditions mimic the physical conditions of high-mass hot cores, in particular the surface temperature of 20~K. Higher temperatures increase the mobility of radicals and result in loss of H atoms by desorption or H$_{2}$ formation, thus limiting the back reaction after a hydrogen bond dissociation. The experiments thus hint that relatively warm conditions on interstellar dust grains are favourable for CH$_{3}$NH$_{2}$ formation.

\section{Discussion}
\label{sec.dis}

\begin{figure*}
\begin{center}
\includegraphics[width=\hsize]{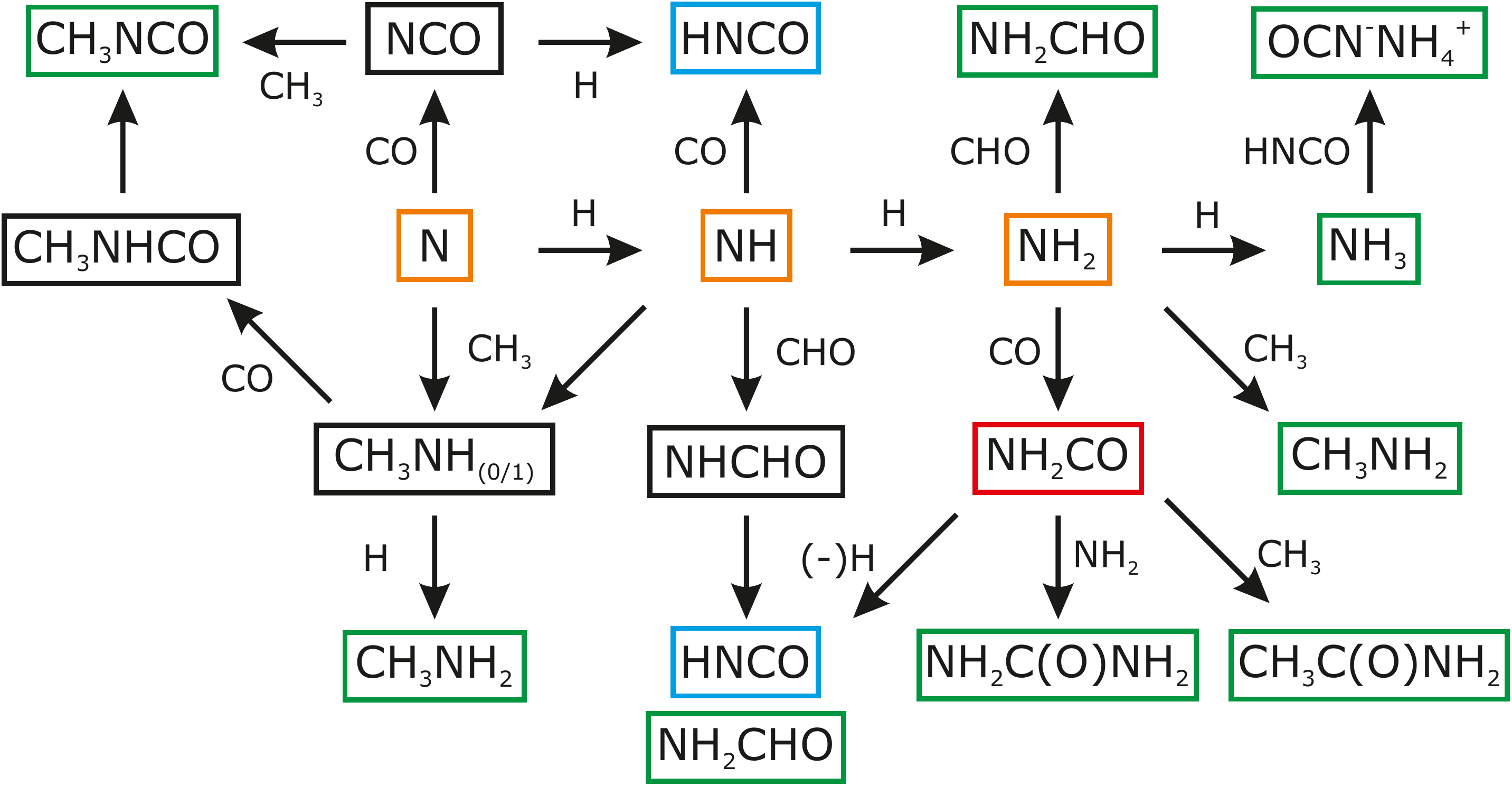}
\caption{Proposed solid-state reaction scheme for the formation of the smallest generation of amides and CH$_{3}$NH$_{2}$, based on the findings of this work. The reactions derive from the nitrogen hydrogenation back bone (orange boxes). Green boxes indicate the products detected in the experiments, while the experimental precursor species HNCO is given in a blue box. CH$_{3}$N$_{0/1}$ indicates an intermediary product that is either the CH$_{3}$N or CH$_{3}$NH radical.}
\label{fig.network}
\end{center}
\end{figure*}

Based on the experimental and observational data obtained in the previous sections, possible reaction pathways to the various products can be identified. For the amides it seems unlikely that hydrogenation reactions with energetic H-atoms of HNCO lead to NH$_{2}$CHO or intermediate radicals (i.e. NH$_{2}$CO and NHCHO). The formation of the amides is therefore explained by non-energetic radical recombination reactions of NH$_{2}$. Energetic radicals, produced by far-UV dissociation of molecules, are not a prerequisite, as inferred from the CO matrix experiments, where these radicals will quickly thermalise.
\begin{equation}
	\rm
	NH_{2} + HCO \rightarrow NH_{2}CHO
	\label{eq.1}
\end{equation}
\begin{equation}
	\rm
	NH_{2} + CO \rightarrow NH_{2}CO
\end{equation}
\begin{equation}
	\rm
	NH_{2}CO + CH_{3} \rightarrow CH_{3}C(O)NH_{2} 
\end{equation}
\begin{equation}
	\rm
	NH_{2}CO + NH_{2} \rightarrow NH_{2}C(O)NH_{2} 
\end{equation}
Equation~\ref{eq.1} is a formation pathway identified by \citet{jones2011}. For the formation of larger amides the carbamoyl radical (NH$_{2}$CO) could be an important intermediate, as was suggested by \citet{agarwal1985}. However, the formation of CH$_{3}$C(O)NH$_{2}$ via CH$_{3}$ + CO $\rightarrow$ CH$_{3}$CO followed by addition of the NH$_{2}$ radical can not be ruled out, but is less likely due to the non-detection of acetaldehyde (CH$_{3}$CHO) in these experiments. 

The non-detection of CH$_{3}$NHCHO has various implications for chemical pathways. It indicates that hydrogenation of CH$_{3}$NCO, an abundantly formed molecule in the ices investigated in this work, does not result in the formation of CH$_{3}$NHCHO. Hydrogenation experiments of pure CH$_{3}$NCO have to be performed to confirm this. For other pathways, such as CH$_{3}$ + NHCHO or CH$_{3}$ + CHO radical addition, the required radicals seem not to be formed in sufficient amounts, easily destroyed or consumed to form other products. If HNCHO is formed in a similar radical addition reaction as NH$_{2}$CO (i.e. NH + CHO), reduction or hydrogenation reactions can result in HNCO and NH$_{2}$CHO formation. Since these reactions are likely to occur for the NH$_{2}$CO radical as well, this does raise the question how NHCHO and NH$_{2}$CO differ in terms of stability and reactivity. Further research into the behaviour of these radicals is of interest to understand the formation of amides in interstellar ices. Analysis of a TPD trace of $m/z$ 59 in \citet[][their Fig. 3]{bossa2012} shows that peak desorption occurs just under 200~K. This is between the peak desorption temperatures of CH$_{3}$NHCHO and CH$_{3}$C(O)NH$_{2}$, but could mean that  CH$_{3}$NHCHO is indeed formed and co-desorbs with NH$_{2}$CHO. Therefore, an ice formation route of CH$_{3}$NHCHO via CH$_{3}$NH + CHO seems possible, but inefficient given its non-detection in this work.

The presence of CH$_{3}$NH$_{2}$ is best explained by the radical addition reaction CH$_{3}$+NH$_{2}$, a reaction pathway proposed by \citet{garrod2008}. However, reactions of CH$_{2}$ or NH radicals followed by hydrogenation can not be ruled out:
\begin{equation}
	\rm
	CH_{2}  + NH \rightarrow CH_{2}NH \ \xrightarrow[\text{}]{\text{+H}} CH_{3}NH_{2}
\end{equation}
\begin{equation}
	\rm
	CH_{3}  + NH \rightarrow CH_{3}NH \ \xrightarrow[\text{}]{\text{+H}} CH_{3}NH_{2}
\end{equation}
\begin{equation}
	\rm
	CH_{2} + NH_{2} \rightarrow CH_{2}NH_{2} \ \xrightarrow[\text{}]{\text{+H}} CH_{3}NH_{2}
\end{equation}
The formation of an intermediate radical is of interest for the formation of CH$_{3}$NHCO, as mentioned earlier, but it also opens up another pathway for CH$_{3}$NCO formation. \citet{cernicharo2016} and \citet{ligterink2017a} suggest the solid-state reaction CH$_{3}$ + NCO, but an alternative could be:
\begin{equation}
	\rm
	CH_{3}NH  + CO \rightarrow CH_{3}NHCO \ \xrightarrow[\text{}]{\text{-H}} CH_{3}NCO
\end{equation}
which shows some similarity with the formation mechanism via a HCN$^{...}$CO van der Waals complex in combination with hydrogenation, proposed by \citet{majumdar2018} to be the dominant pathway to form CH$_{3}$NCO in interstellar ice mantles. 

The identified experimental reaction pathways lead to the interstellar ice reaction scheme presented in Fig.~\ref{fig.network}. Starting from atomic nitrogen, NH, NH$_{2}$ and NH$_{3}$, a variety of reactions, mainly with CO and CHO, form HNCO, CH$_{3}$NCO, NH$_{2}$CHO, CH$_{3}$C(O)NH$_{2}$ and NH$_{2}$C(O)NH$_{2}$. CH$_{3}$NH$_{2}$ is formed from CH$_{x}$ + NH$_{y}$ radical addition reactions, followed by hydrogenation if necessary. In the experiment, far-UV radiation is used to produce the radicals from CH$_{4}$ and HNCO. On interstellar dust grains these radicals can also be formed in hydrogenation sequences of carbon and nitrogen atoms. In interstellar ice, HNCO is likely not at the basis of the formation of amides, but rather formed simultaneously. 

Whether complex molecules form in the ice mantles on interstellar dust grains or in gas-phase reactions is still an ongoing debate. In this work it was possible to directly compare the ratio of [CH$_{3}$C(O)NH$_{2}$] / [NH$_{2}$CHO] in laboratory experiments and observational data, and indicate that their interstellar gas-phase presence, particularly in hot cores, can be explained by formation in interstellar ice mantles and subsequent thermal desorption. However, for these molecules gas-phase formation pathways have been proposed, such as NH$_{2}$ + H$_{2}$CO $\rightarrow$ NH$_{2}$CHO + H \citep{barone2015,skouteris2017}. To strengthen the claim that solid-state reactions form amide bearing molecules, more comparisons between observational and laboratory TPD data of the molecules in this network are needed. Detections of these molecules, preferably towards comparatively less complex low-mass YSOs, will aid this process by expanding observational statistics. Experimentally, TPD data on the species that are formed in different solid-state experiments is needed. In the present experiments HNCO is used as a precursor and therefore ratios with respect to this molecule can not be determined. Energetic processing of, for example, CH$_{4}$:CO:NH$_{3}$ ice mixtures can be employed to circumvent this problem. 

\section{Conclusions}
\label{sec.con}

This work has investigated the solid-state formation of amides and amines in far-UV irradiated CH$_{4}$:HNCO ice mixtures. Observations towards the low-mass sun-like protostar IRAS~16293--2422B were analysed in search of CH$_{3}$NHCHO and CH$_{3}$C(O)NH$_{2}$. The conclusions of this work are summarised as follows:

\begin{itemize}
\item Acetamide (CH$_{3}$C(O)NH$_{2}$), formamide (NH$_{2}$CHO), carbamide (NH$_{2}$CONH$_{2}$), methyl isocyanate (CH$_{3}$NCO) and methylamine (CH$_{3}$NH$_{2}$) are shown to simultaneously form in the solid-state, under conditions mimicking those of ice mantels on interstellar dust grains.  
\item The carbamoyl radical (NH$_{2}$CO) is identified as an important reaction intermediate in the formation of amides, specifically CH$_{3}$C(O)NH$_{2}$ and NH$_{2}$C(O)NH$_{2}$.
\item CH$_{3}$C(O)NH$_{2}$ is tentatively identified in ALMA observations towards the low-mass sun-like protostar IRAS~16293--2422B at column densities of 9$\times$10$^{14}$ cm$^{-2}$ and 25$\times$10$^{14}$ cm$^{-2}$ for $T_{\rm ex}$ = 100 and 300~K, respectively. The resulting ratio of [CH$_{3}$C(O)NH$_{2}$] / [NH$_{2}$CHO] = 0.09--0.25 matches with abundances found towards high-mass YSOs. CH$_{3}$NHCHO is not detected down to an upper limit column density of $\leq$1$\times$10$^{15}$ cm$^{-2}$.
\item A comparison between experimental TPD data and observations of CH$_{3}$C(O)NH$_{2}$ and NH$_{2}$CHO, indicates that solid-state chemistry on interstellar dust grains likely is responsible for the formation of these two species. 
\item A solid-state reaction network for the smallest generations of amides and amide-like molecules is proposed, focussing on reactions between N, NH and NH$_{2}$ radicals with CO and CHO, followed by CH$_{3}$ addition.
\end{itemize}

\section*{Acknowledgements}

The authors thank A. Belloche and V. Ilyushin for sharing the rotational spectroscopy files of acetamide and N-methylformamide and for useful discussions on the subject of amide formation. This paper makes use of the following ALMA data: ADS/JAO.ALMA\#2012.1.00712.S, ADS/JAO.ALMA\#2013.1.00278.S and ADS/JAO.ALMA\#2016.1.01150.S. ALMA is a partnership of ESO (representing its member states), NSF (USA) and NINS (Japan), together with NRC (Canada), NSC and ASIAA (Taiwan), and KASI (Republic of Korea), in cooperation with the Republic of Chile. The Joint ALMA Observatory is operated by ESO, AUI/NRAO and NAOJ. NL is funded by the Origins Centre. Astrochemistry in Leiden is supported by the European Union A-ERC grant 291141 CHEMPLAN, by the Netherlands Research School for Astronomy (NOVA) and by a Royal Netherlands Academy of Arts and Sciences (KNAW) professor prize. CryoPAD2 was realized with NOVA and NWO (Netherlands Organisation for Scientific Research) grants, including a NWO-M grant, a NWO-Vici grant and funding through the Dutch Astrochemistry Network II. JKJ acknowledges support from the European Research Council (ERC) under the European Union's Horizon 2020 research and innovation programme (grant agreement No$\sim$646908) through ERC Consolidator Grant ``S4F''.




\bibliographystyle{mnras}
\bibliography{lib} 


\appendix

\section{Additional mass spectrometric reference data}
\label{ap.TPD_for}

\begin{figure}
\begin{center}
\includegraphics[width=\hsize]{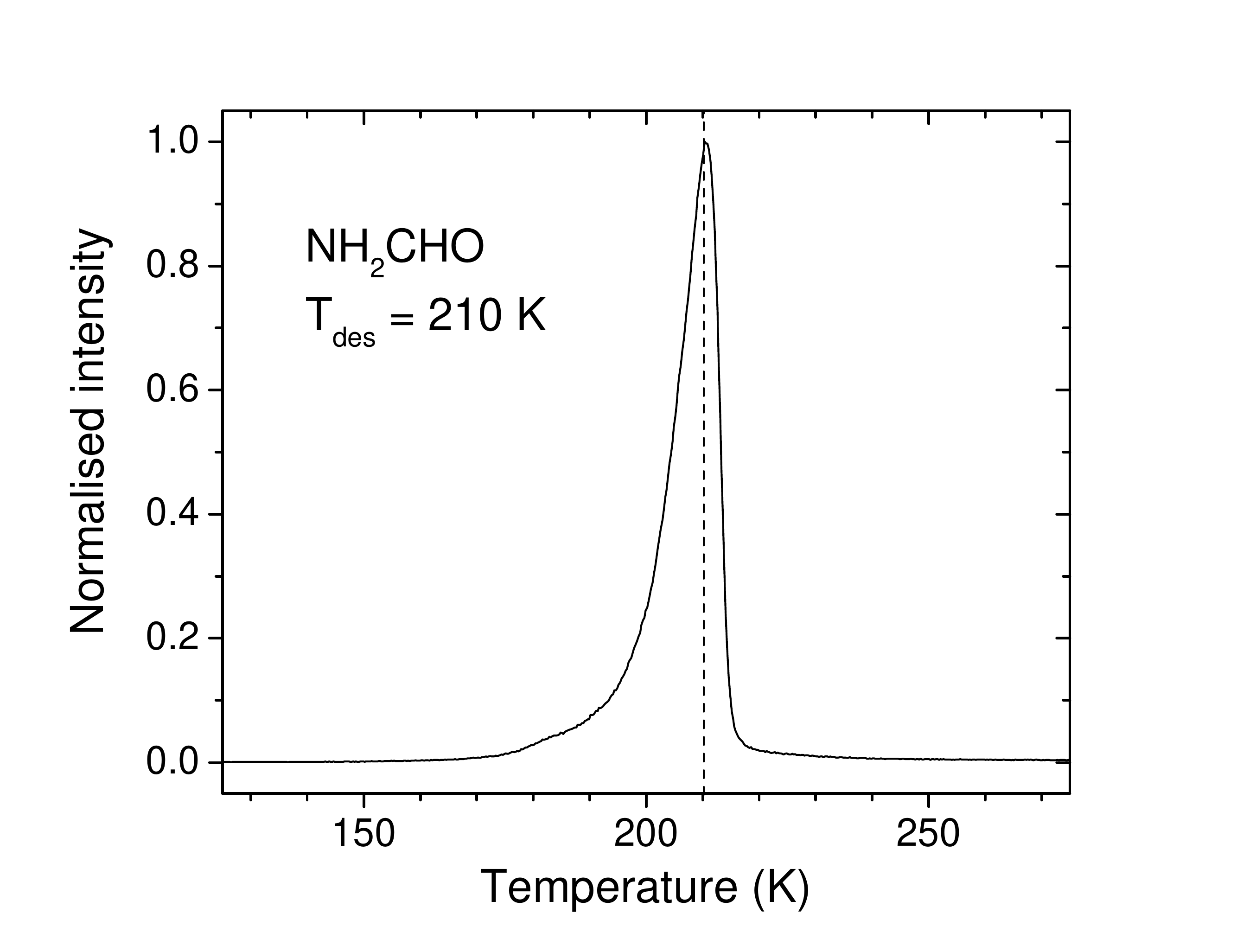}
\caption{TPD trace of $m/z$ 45 of NH$_{2}$CHO.}
\label{fig.for_TPD}
\end{center}
\end{figure}

\begin{figure}
\begin{center}
\includegraphics[width=\hsize]{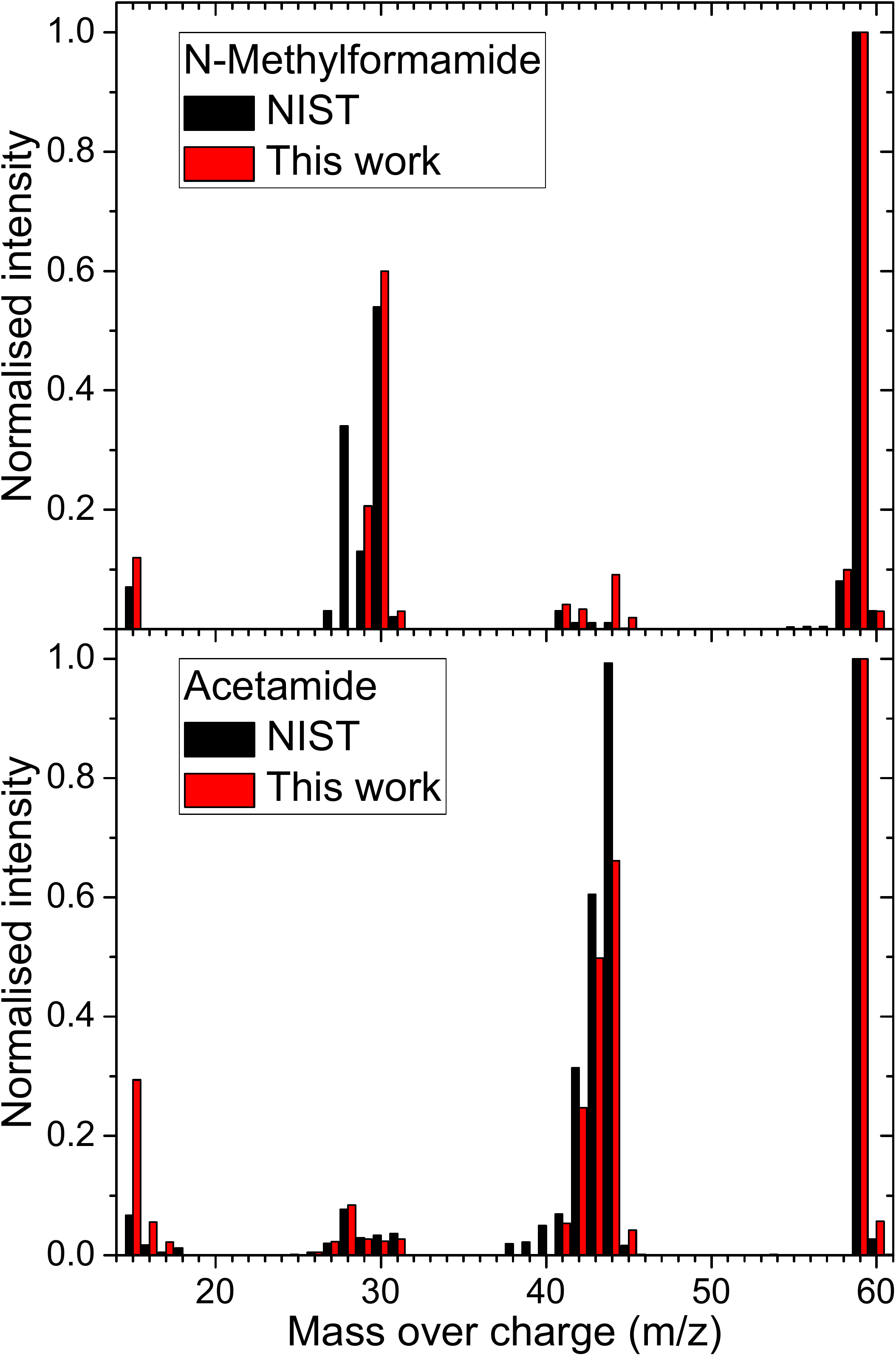}
\caption{Fragmentation pattern comparison between NIST data (black) and values measured in this work for N-methylformamide (CH$_{3}$NHCHO, top in red) and acetamide (CH$_{3}$C(O)NH$_{2}$, bottom in red).}
\label{fig.NISTcomp}
\end{center}
\end{figure}

Figure \ref{fig.for_TPD} shows the TPD trace of formamide (NH$_{2}$CHO) between 150 and 300~K. The peak desorption temperature is found at 210~K. Figure \ref{fig.NISTcomp} shows a comparison of the N-methylformamide (CH$_{3}$NHCHO) and acetamide (CH$_{3}$C(O)NH$_{2}$) fragmentation pattern between $in\ situ$ measurements on CryoPAD2 and reference data from the NIST database, both normalised to the main mass $m/z$ 59. Mainly small deviations in the pattern intensity are seen, with the notable exception of the $m/z$ 44 and 17 fragments of CH$_{3}$C(O)NH$_{2}$ which show large discrepancies compared to the NIST data. Throughout this work the $in\ situ$ measured fragmentation patterns of CH$_{3}$NHCHO and CH$_{3}$C(O)NH$_{2}$ will be used. 

\begin{table*}
     \caption[]{Mass fragmentation patterns of NH$_{3}$, HNCO, NH$_{2}$CHO, CH$_{3}$C(O)NH$_{2}$ and CH$_{3}$NHCHO at selected masses.}
         $$
         \begin{tabular}{l l l l l l l}
            \hline
            \hline
            \noalign{\smallskip}
            $m/z$ & NH$_{3}$ & HNCO$^{a}$ & CH$_{3}$NH$_{2}$ & NH$_{2}$CHO & CH$_{3}$C(O)NH$_{2}$~$^{b}$ & CH$_{3}$NHCHO$^{b}$ \\
            \noalign{\smallskip}
            \hline
            \noalign{\smallskip}
            15 & 0.08 & 0.07 & 0.04 & 0.01 & 0.29 & 0.12 \\
            16 & 0.80 & -- & -- & 0.12 & 0.06 & -- \\
            17 & 1.00 & -- & 0.01 & 0.33 & 0.02 & -- \\
            27 & -- & 0.02 & 0.08 & 0.08 & 0.02 & -- \\
            28 & -- & 0.07 & 0.54 & 0.06 & 0.08 & -- \\
            29 & -- & 0.14 & 0.12 & 0.34 & 0.03 & 0.21 \\
            30 & -- & \textcolor{blue}{0.02} & \textcolor{red}{1.00} & \textcolor{blue}{0.01} & \textcolor{blue}{0.02} & \textcolor{blue}{0.21}\\
            31 & -- & -- & \textcolor{red}{0.66} & -- & \textcolor{blue}{0.03} & \textcolor{blue}{0.03}\\
            42 & -- & 0.22 & -- & 0.02 & 0.25 & 0.03 \\
            43 & -- & 1.00 & -- & 0.11 & 0.50 & -- \\
            44 & -- & 0.02 & -- & 0.25 & 0.66 & 0.09 \\
            45 & -- & -- & -- & 1.00 & 0.04 & 0.02 \\
            59 & -- & -- & -- & -- & 1.00 & 1.00 \\          
            \noalign{\smallskip}
            \hline
            
         \end{tabular}     
         $$     
		\emph{\rm Notes. $^{a}$Fragmentation pattern from \citet{bogan1971}, $^{b}$Fragmentation pattern measured in this work. Other fragmentation patterns are adapted from the NIST database.}
	\label{tab.frag_comp}
\end{table*}

\begin{figure}
\begin{center}
\includegraphics[width=\hsize]{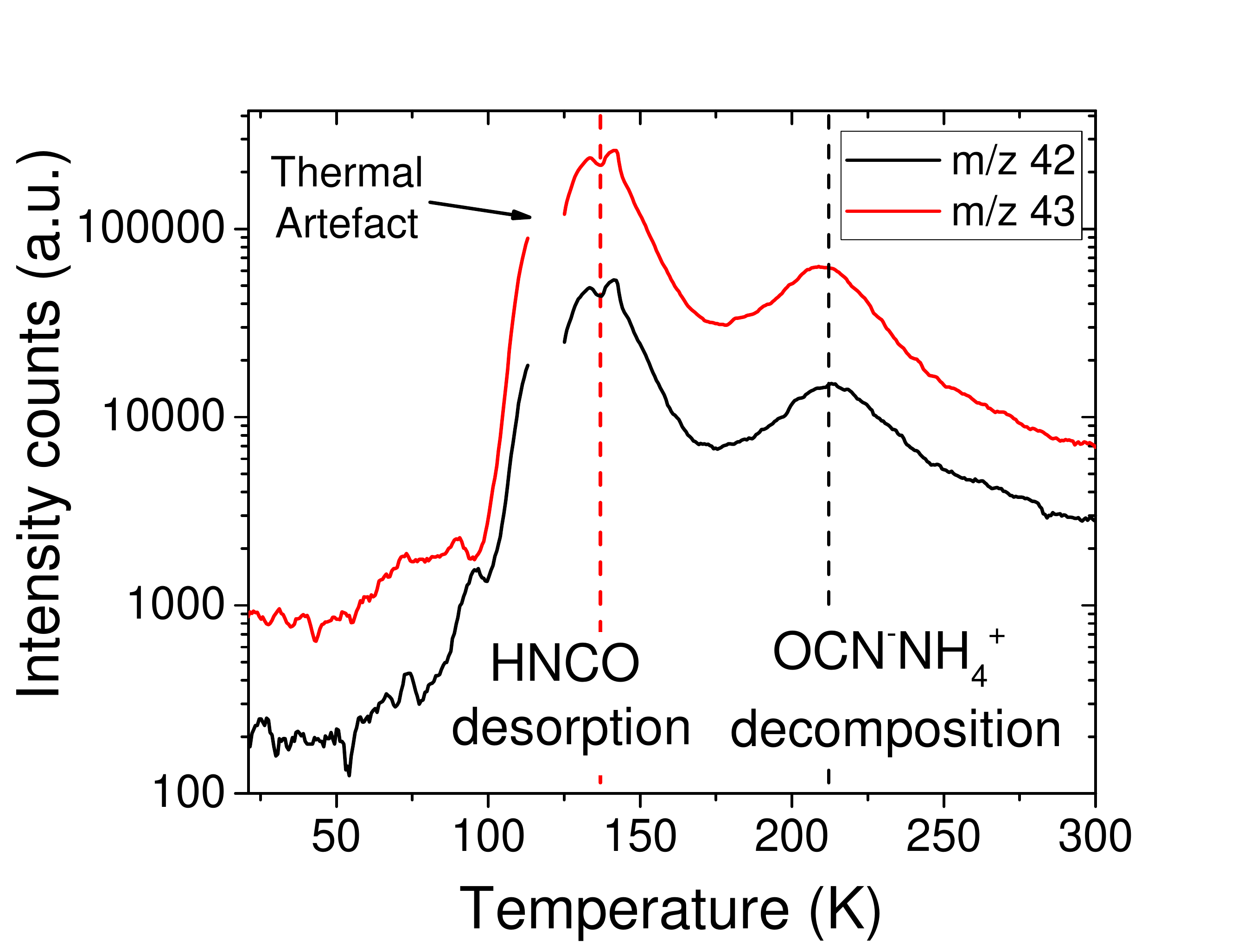}
\caption{TPD trace of $m/z$ 43 and 42 of HNCO.}
\label{fig.full}
\end{center}
\end{figure}

Figure \ref{fig.full} shows the TPD trace of $m/z$ 43 and 42 over the full temperature range of 20 to 300~K. The data from Exp. 1, also displayed in Fig. \ref{fig.primary_amides}, are used. The data shows the main thermal desorption peak of HNCO at $\sim$130~K, after which HNCO gas present in the vacuum chamber is being pumped away, resulting in a trailing slope. At $\sim$210~K, the $m/z$ 43 and 42 increase again, resulting in a small peak on top of the trailing slope. This increase is the result of the thermal decomposition of OCN$^{-}$NH$_{4}^{+}$ and OCN$^{-}$CH$_{3}$NH$_{3}^{+}$ and subsequent release of HNCO to the gas-phase.

\section{Observed rotational transitions of acetamide}
\label{ap.spec}

\begin{table*}
     \caption[]{Identified CH$_{3}$C(O)NH$_{2}$ lines towards IRAS 16293--2422B}
         $$
         \begin{tabular}{l l l l l l}
            \hline
            \hline
            \noalign{\smallskip}
            Upper state & Lower state & Frequency & $E_{\rm up}$ & $A_{\rm ij}$ & Line blending\\
            $J'$ $K'_{a}$ $K'_{b}$ & $J''$ $K''_{a}$ $K''_{b}$ & (MHz) & (K)		& ($\times$10$^{-5}$ s$^{-1}$) & \\
            \noalign{\smallskip}
            \hline
            \noalign{\smallskip}
            16 6 10 & 15 7 9 & 225 150.6 & 112.5 & 44.9 & HCOCH$_{2}$OH \\
            \noalign{\smallskip}
			16 7 10 & 15 6 9 & 225 218.9 & 112.5 & 45.0 & unblended \\
			\noalign{\smallskip}
			20 4 17 & 19 4 16 & 236 203.3 & 140.0 & $\phantom{0}$5.2 & gGg' (CH$_{2}$OH)$_{2}$ \\ 
			20 3 17 & 19 3 16 & 236 203.3 & 140.0 & $\phantom{0}$5.2 & gGg' (CH$_{2}$OH)$_{2}$ \\ 
			20 3 17 & 19 4 16 & 236 203.3 & 140.0 & 77.0 & gGg' (CH$_{2}$OH)$_{2}$ \\ 
			20 4 17 & 19 3 16 & 236 203.3 & 140.0 & 77.0 & gGg' (CH$_{2}$OH)$_{2}$ \\ 
			\noalign{\smallskip}
			21 2 19 & 20 3 18 & 239 780.3 & 143.9 & $\phantom{0}$1.5 & unblended\\
 			21 2 19 & 20 2 18 & 239 780.3 & 143.9 & 90.4 & unblended\\
	 		21 3 19 & 20 2 18 & 239 780.3 & 143.9 & $\phantom{0}$1.5 & unblended\\
	 		21 3 19 & 20 3 18 & 239 780.3 & 143.9 & 90.4 & unblended\\
            \noalign{\smallskip}

            \hline
            
         \end{tabular}    
         $$     
	\emph{\rm Notes. Parameters adapted from \citet{ilyushin2004}.} 
	\label{tab.transitions}
\end{table*}


\bsp	
\label{lastpage}
\end{document}